\documentclass[prb,final,twocolumn,amsmath,eqsecnum,amssymb,superscriptaddress,showpacs]{revtex4-2}
\usepackage{graphicx,graphics}
\usepackage{color}

\usepackage{epsfig}
\usepackage{amsmath}
\usepackage{amssymb}
\usepackage{bm}

\begin{document}

\title{Spin-orbital mechanisms for negative thermal expansion
       in Ca$_2$RuO$_4$}

\author{Wojciech Brzezicki}
\affiliation{\mbox{Institute of Theoretical Physics,
Jagiellonian University,
Prof. Stanis\l{}awa \L{}ojasiewicza 11, PL-30348 Krak\'ow, Poland}}

\author{Filomena Forte}
\affiliation{Consiglio Nazionale delle Ricerche, CNR-SPIN,
             IT-84084 Fisciano (SA), Italy}
\affiliation{Dipartimento di Fisica ``E. R. Caianiello", Universit\`a di Salerno,
             Via Giovanni Paolo II 132, IT-84084 Fisciano (SA), Italy }

\author{Canio Noce}
\affiliation{Dipartimento di Fisica ``E. R. Caianiello", Universit\`a di Salerno,
             Via Giovanni Paolo II 132, IT-84084 Fisciano (SA), Italy }

\author{Mario Cuoco}
\affiliation{Consiglio Nazionale delle Ricerche, CNR-SPIN,
             IT-84084 Fisciano (SA), Italy}
\affiliation{Dipartimento di Fisica ``E. R. Caianiello", Universit\`a di Salerno,
             Via Giovanni Paolo II 132, IT-84084 Fisciano (SA), Italy }

\author{Andrzej M. Ole\'s$\,$}
\affiliation{Max Planck Institute for Solid State Research,
             Heisenbergstrasse 1, D-70569 Stuttgart, Germany }
\affiliation{\mbox{Institute of Theoretical Physics,
Jagiellonian University,
Prof. Stanis\l{}awa \L{}ojasiewicza 11, PL-30348 Krak\'ow, Poland}}

\date{\today} 

\begin{abstract}
The phenomenon of negative thermal expansion (NTE) deals with the increase of the lattice parameters and the volume of the
unit cell when the material is thermally cooled. The NTE is typically
associated with thermal phonons and anomalous spin-lattice coupling at
low temperatures. However, the underlying mechanisms in the presence of
strong electron correlations in multi-orbital systems are not yet fully
established. Here, we investigate the role of Coulomb interaction in the
presence of lattice distortions in setting out the NTE effect, by focusing
on the physical case of layered Ca$_2$RuO$_4$ with $d^4$ configuration
at each Ru ion site. We employ the Slater-Koster parametrization to
describe the electron-lattice coupling through the dependence of the
$d-p$ hybridization on the Ru-O-Ru bond angle. The evaluation of the
minimum of the free energy at finite temperature by fully solving the
multi-orbital many-body problem on finite size cluster allows us to
identify the regime for which the system is prone to exhibit NTE effects.
The analysis shows that the nature of the spin-orbital correlations is
relevant to drive the reduction of the bond angle by cooling, and in
turn the tendency toward a NTE. This is confirmed by the fact that a
changeover of the electronic and orbital configuration from $d^4$ to
$d^3$ by transition metal substitution is shown to favor the occurrence
of NTE in Ca$_2$RuO$_4$. This finding is in agreement with the
experimental observations of a NTE effect which is significantly
dependent on the transition metal substitution in the Ca$_2$RuO$_4$
compound.
\end{abstract}

\maketitle

\section{Introduction}

Nowadays it is widely accepted that strong electron correlations play a
prominent role in achieving and controlling a large variety of quantum
phases, physical phenomena and effects related to cooperative behaviors
that cannot be described from the properties of individual electrons.
Emblematic cases are the unconventional pairing in high $T_c$
superconductors \cite{Keimer15}, as well as metal-insulator transitions
\cite{Imada98} or emergent excitations such as monopoles
\cite{Castelnovo08} and skyrmions \cite{Mul09}. Transition metal oxides
(TMOs) are paradigmatic quantum materials \cite{Evans21} in this context,
as they are marked by strong correlations that result in several types
of electronic phases, including charge- and spin-ordered states.
Prominent examples are found among the $3d$ TMOs, where unconventional
transport properties, ordering phenomena, and unusual spectroscopic
properties are observed. It was argued that the comparably weak spatial
extension of $3d$ orbitals may lead to large electronic Coulomb
interactions, competing with kinetic contributions. Depending on
crystal-field, hybridization, Hund's exchange, and band filling, this
interplay can lead to renormalized metallic behavior or induce Mott
insulating behavior. In $4d$ and $5d$ oxides, spin-orbit coupling acts
on an energy scale comparable to the other energy scales of the system,
and the observed electronic state is the result of a complex interplay
of Coulomb interactions, spin–orbit splitting, and crystal field effects.

Interestingly, when the electron correlations are somewhat weaker, as in
$4d$ or $5d$ TMOs, other challenging effects may arise due to the subtle
competition between electronic and lattice degrees of freedom. One of
most extraordinary one is the negative thermal expansion (NTE) in
Ca$_2$RuO$_4$-derived layered materials. The phenomenon of NTE
\cite{Azuma15,Takenaka18,Chen15} deals with the observation of a
shrinking of the lattice parameters when the material is heated or
\textit{vice versa} with a lattice expansion by thermal cooling.

Although the phenomenon is not broadly observed, it has a remarkable
impact for several applications in electronics, optics, and for the
design of thermal engines or medical products \cite{Takenaka12,Lia21}.
The origin of NTE is quite intricate as, apart from the structural
modifications, it can involve the coupling of electrons, spin, orbital
degrees of freedom. In materials such as ZrW$_2$O$_8$ and ZnCr$_2$Se$_4$,
the NTE is mainly due to oxygen vibrational modes
\mbox{\cite{Cao08,Martineck68,Mary96,Ernst98,Hua16}} and anharmonicity
related to spin canting and spin-lattice coupling. Hence, while the
manifestation of the effect is directly provided by the modification of
the lattice parameters, a fundamental question arises on whether the
electronic correlations, and the resulting spin-orbital couplings, would
cooperate or compete with the tendency of the lattice to exhibit a NTE.
This issue is particularly relevant in materials with Mott physics and
nontrivial magnetic ordering, with electronic degrees of freedom
being dominated by local Coulomb interaction, especially in the presence
of multi-orbital configurations having significant spin-orbital
entanglement.

Looking more inside at the Ca$_2$RuO$_4$ family, it has been reported that a
tiny percentage of Cr substituting Ru produces a 1$\%$ volume reduction
that has been ascribed to the collapse of the orbital and magnetic
ordering \cite{Qi10}. Remarkably, substituting Ru with $M$ ion ($M=$Cr,
Mn, Fe, or Cu) generally yields NTE effects in Ca$_2$Ru$_{1-x}M_x$O$_4$
\cite{Tak17,Koo21,Qi12}, with the onset following the behavior of the
metal-insulator and magnetic ordering transition temperatures
\cite{Qi12}. Therefore, the doping by transition metal (TM) ions plays
an important role in unveiling a generic tendency toward the phenomenon
of NTE and altering the strength of the spin-orbital interactions while
keeping the system in the insulating regime \cite{Brz15,Brz18,Brz20}.

It is thus compelling to attribute the observed NTE to a mechanism where
electronic correlations play a critical role. Insulating Ca$_2$RuO$_4$
is a representative case of a multi-orbital correlated TMO, where both
spins and orbitals develop their own dynamics and are coupled to each
other through lattice distortions. In this frame, the mechanism of
NTE can thus contain a high degree of complexity. In TMOs, the distance
of the TM ions is related to the TM$-$O$-$TM bond angle which in turn
is determined by the rotation of the TM$-$O$_6$ octahedra around a
given crystallographic axis. Then, a reduction of the lattice parameters
can be obtained by a deviation from 180$^{\circ}$ of the TM$-$O$-$TM
bond angle (see Fig. \ref{fig:1}). While the reduction of the
TM$-$O$-$TM angle is expected to reduce the kinetic
energy in an insulator described by a single orbital, the energy balance
for multi-orbital electronic systems is not obvious, especially in the
presence of Coulomb interaction.

Therefore, various cooperative effects may materialize. The electron
localization reduces the kinetic energy of the electrons and,
simultaneously, the lattice expansion may cost energy due to
electron-lattice interaction. Hence, the lattice expansion may influence
the kinetic energy variation even further. Moreover, the spin-lattice
coupling also plays an important role, particularly when strong
competition between ferromagnetic (FM) and antiferromagnetic (AF)
interaction occurs since this interplay often leads to a strong bond
frustration or lattice anomaly. Interestingly, the frustration may be
released in a phase transition by which a certain type of magnetic
order is stabilized.

Such competing mechanisms depend on the TM$-$O$-$TM bond angle, which
directly enters in setting out the most favorable structural
configurations. In this context, the short-range orbital or magnetic
or spin-orbital correlations play a crucial role and decide about the
energy gain or loss when a variation of the lattice parameters occurs.

Taking into account these multiple components entering into the
phenomenology of the NTE, we aim to investigate such physical scenario
by assessing the role of Coulomb interaction and spin-orbital
correlations in setting out the thermal evolution of the TM$-$O$-$TM
lattice configuration. In order to determine the impact of the
TM$-$O$-$TM configuration in the electron correlated configuration, we
employ the Slater-Koster parametrization to describe the
electron-lattice coupling through the dependence of the $d-p$
hybridization on the TM$-$O$-$TM bond
angle. The evaluation of the minimum of the free energy at finite
temperature is achieved by a complete solution of the multi-orbital
many-body problems on finite-size clusters. This approach allows us to
identify the electronic regime where the system can exhibit NTE effects.
The analysis shows that the nature of spin-orbital correlations is
relevant to drive the reduction of the Ru$-$O$-$Ru bond angle by thermal
cooling, and in turn the tendency to develop a NTE. Indeed, the
changeover of the electronic and orbital configuration from $d^4$ to
$d^3$ by TM substitution clearly supports the role of
electron correlations to account for the occurrence of NTE in
Ca$_2$RuO$_4$. This finding agrees with the experimental observation
of a NTE effect, which exhibits significant dependence on the TM
substitution in the Ca$_2$RuO$_4$ compound.

The paper is organized as follows. First, in Sec. \ref{sec:bond} we
consider a single TM$-$O$-$TM bond with two $d^4$ ions and investigate
whether this bond may generate a configuration compatible with the
occurrence of a NTE effect. Second, we show that the TM$-$O$-$TM bond
with one $d^3$ ion substituting the $d^4$ ion has a tendency to favor
bond distortions that in general support the NTE in a planar perovskite.
The search for an optimal bond angle configuration in the presence of
Coulomb interaction at the TM site is further analyzed with a
$2\times 2$ plaquette. We derive an effective low-energy model that
properly includes all the $d-p$ charge transfer processes, the bond
angles, and the local Coulomb interaction. In Sec. \ref{sec:44} we
investigate the case with $d^4$ charge configurations while in Sec.
\ref{sec:34} we consider the case of a single $d^3$ impurity replacing
a $d^4$ site. The study of the planar plaquette allows us to address
the role of orbital correlations in setting out electronic states which
are compatible with the NTE effects. In particular, orbital
configurations or electronic mechanisms that break the rotation by
90$^\circ$ in the square lattice are particularly relevant for favoring
lattice distortions that support the NTE effects in Ca$_2$RuO$_4$.
Our main conclusions are summarized in Sec. \ref{sec:summa}. The
superexchange Hamiltonians are explicitly derived and reported together
with the complete set of their coefficients in the Appendices
\ref{sec:app44} and \ref{sec:app34}.
In the Appendix \ref{sec:app_renorm} we provide details about the evolution of the optimal angle as a function of the Hund's coupling and of the spin-orbit interaction. In the Appendix \ref{sec:1bond_eff} we present the results of the effective spin-orbital model for a single bond with $d^4-d^4$ and $d^3-d^4$ configurations. Finally, in the Appendix \ref{sec:latt} we introduce a phenomenological electron-lattice potential
to describe the lattice feedback on the electronic system that is able to limit the bond angle within a physical range.


\section{Single TM$-$O$-$TM bond analysis}
\label{sec:bond}

\subsection{The model}

\begin{figure}[t!]
\includegraphics[width=0.48\textwidth]{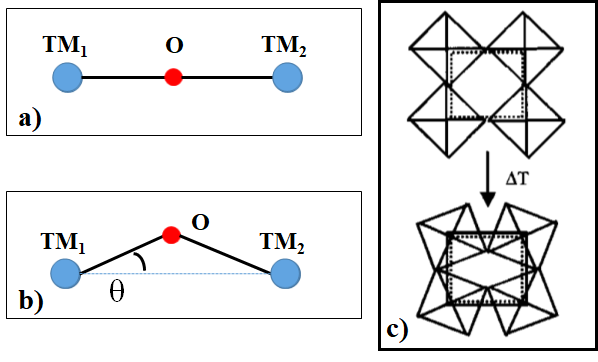}
\caption{Schematic representation of the TM$_1-$O$-$TM$_2$ bond
for the case of:
(a) an undistorted configuration with $\theta=0$,
(b) a non-vanishing bond angle due to the misalignment between the
TM$_1-$O and TM$_1-$TM$_2$ directions.
In (c) we provide a sketch to depict how a variation of the bond angle
for a square geometry of the TM$-$O lattice can lead to a modification
of the unit cell area. In particular, a thermal gradient $\Delta T$ that
induces an increase of the TM$-$O$-$TM bond angle would lead to a
reduction of the unit cell.}
\label{fig:1}
\end{figure}

In this Section, we investigate the energetically most favorable
TM$-$O$-$TM bond angle configuration, which is realized when Coulomb
interaction at the TM element is fully taken into account. To this end,
we consider a multi-orbital $p-d$ electronic system suitable for oxide
materials. This includes all the atomic terms arising from the local
Coulomb interaction and the orbital-dependent connectivity between a TM
element and the oxygen ligands that are allowed in the planar square
lattice geometry, typical for tetragonal perovskites. First, we consider
the case of a single bond made of two TM elements, i.e., TM$_1$ and
TM$_2$, linked via an oxygen (O) ion between them, as schematically
depicted in Fig. \ref{fig:1}.

Note that the deviation of the bond angle away from 180$^{\circ}$ is
directly related to the angle $\theta$, formed by the
TM$-$O distance and the crystal axis. Hence, in the following, we will
refer to $\theta$ as a measure of the bond angle, with $\theta$=0
corresponding to an undistorted 180$^{\circ}$ bond, and the maximal
$\theta=45^{\circ}$ corresponding to a 90$^{\circ}$ TM$-$O$-$TM bond.
The fundamental physical motivation for this study is to assess
a role that can be played by the electron-electron correlations in
setting the bond angle between TM elements and including their
effects on the thermal evolution of the bond angle.

Within this framework, we aim to evaluate the role of spin, orbital and
charge-transfer processes in setting out the energetically most
favorable TM$-$O$-$TM angle. In particular, we consider how the Coulomb
interaction at the TM site leads to the optimal TM$-$O$-$TM bond angle
and whether the resulting configuration is related to the spin-orbital
correlations or other electronic parameters associated with the
tetragonal distortions and spin-orbit coupling. The analysis is
performed by solving exactly a model Hamiltonian for the
TM$_1-$O$-$TM$_2$ cluster, with two TM ions at its ends.

To describe the TM$_1-$O$-$TM$_2$ cluster, we employ a
microscopic model Hamiltonian for the bands close to the Fermi level for
the itinerant electrons within the TM$-$O plane by considering the
interaction terms at the TM sites and the kinetic term along the TM$-$O
bond. Since we are focusing here on low-spin configurations for the
$d^4$ electronic states, the local TM-Hamiltonian $H_{\mathrm{loc}}$
\cite{Cuoco06a,Cuoco06b,Forte10} includes the complete Coulomb
interaction projected on the $t_{2g}$ electrons, the spin-orbit
coupling, and the tetragonal crystal field (CF) potential. Therefore,
the $H_{\mathrm{loc}}$ at site $i$ is,
\begin{equation}
H_{\mathrm{loc}}(i)=H_{e-e}(i)+H_{\mathrm{SOC}}(i)+H_{\mathrm{CF}}(i)\, .
\end{equation}

Here, the on-site Coulomb, spin-orbit, and CF terms are expressed at
site $i$ by
\begin{eqnarray}
H_{e-e}(i)&=&
U\sum\limits_{\alpha} n_{i\alpha\uparrow}n_{i\alpha\downarrow}
+J_{H}\sum\limits_{\alpha\ne\beta}
d_{i\alpha\uparrow}^{\dagger}d_{i\alpha\downarrow}^{\dagger}
d_{i\beta\uparrow}^{}d_{i\beta\downarrow}^{},  \nonumber\\
&+&\left(U^{'}-\frac{ J_{\mathrm{H}}}{2}\right)
\sum\limits_{\alpha <\beta }n_{i\alpha }n_{i\beta }
-2J_{\mathrm{H}}\sum\limits_{\alpha<\beta}
({\overrightarrow{S}})_{i\alpha}\!\cdot\!({\overrightarrow{S}})_{i\beta}, \nonumber\\
\\
H_{\mathrm{SOC}}(i)&=&\lambda \sum\limits_{\alpha ,\sigma }
\sum_{\beta ,\sigma^{^{\prime }}}\,                                                                                    d_{i\alpha \sigma }^{\dagger }\,({\overrightarrow{l}})_{\alpha\beta}
\cdot ({\overrightarrow{s}})_{\sigma \sigma ^{^{\prime }}}\,
d_{i\beta\sigma ^{^{\prime }}},  \\
H_{\mathrm{CF}}(i) &=&\varepsilon_{xy} n_{i,xy}+\left(\varepsilon_{xz}
n_{i,xz}+\varepsilon_{yz} n_{i,yz}\right)\, .
\end{eqnarray}
In these expressions, $i$ labels the site and $\{\alpha,\beta\}$ are
indices running over the three orbitals in the $t_{2g}$ sector, i.e.,
\mbox{$\alpha,\beta\in\{d_{xy},d_{xz},d_{yz}\}$,} and
$d_{i\alpha\sigma}^{\dagger}$ is the creation operator of an electron
with spin $\sigma$ at the site $i$ in the orbital $\alpha$. The
interaction is parametrized by the Kanamori parameters: intra-orbital
Coulomb interaction $U$ and Hund's exchange $J_{\mathrm{H}}$.

The strength of the tetragonal distortions is expressed by the
amplitude $\delta$, with $\delta=(\varepsilon_{xy}-\varepsilon_z)$.
We also consider the possibility of having an orthorhombic splitting,
$\delta_{ort}$ of the $\{xz,yz\}$ orbitals by assuming that
$\varepsilon_{yz}=\varepsilon_{z}+\delta_{ort}$ and
$\varepsilon_{zx}=\varepsilon_{z}-\delta_{ort}$.
$U$ is the intra-orbital Coulomb interaction, $J_H$ Hund's exchange,
and $(U^{^{\prime}}-\frac12 J_H)$ sets the strength of the inter-orbital
electron-electron interaction, $J^{^{\prime}}$ is the pair hopping term.
We assume a rotational invariant condition for
the Coulomb amplitudes, so that $U=U^{^{\prime }}+2J_{H}$, and
$J^{^{\prime}}=J_{H}$. The operator $\overrightarrow{l}$ is the
projection of the angular momentum operators to the $t_{2g}$ subspace,
$(l_{k})_{\alpha\beta}=i\epsilon_{k\alpha\beta}$ such as
$\overrightarrow{l}\times\overrightarrow{l}=-i\overrightarrow{l}$ and
$\overrightarrow{s}_i=\frac{1}{2}\overrightarrow{\sigma}_i$ is the spin
operator at site $i$ expressed through the Pauli matrices
$\overrightarrow{\sigma}_i$.

The matrices for the orbital operators in the $t_{2g}$ manifold are:
\begin{eqnarray*}
l_{x} &=&
\begin{bmatrix}
0 & 0 & 0 \\
0 & 0 & i \\
0 & -i & 0%
\end{bmatrix}
\rightarrow
\begin{bmatrix}
d_{yz} \\
d_{xz} \\
d_{xy}%
\end{bmatrix}%
\  \\
\ l_{y} &=&%
\begin{bmatrix}
0 & 0 & -i \\
0 & 0 & 0 \\
i & 0 & 0%
\end{bmatrix}%
\rightarrow
\begin{bmatrix}
d_{yz} \\
d_{xz} \\
d_{xy}%
\end{bmatrix}
\\
l_{z} &=&%
\begin{bmatrix}
0 & i & 0 \\
-i & 0 & 0 \\
0 & 0 & 0%
\end{bmatrix}%
\rightarrow
\begin{bmatrix}
d_{yz} \\
d_{xz} \\
d_{xy}%
\end{bmatrix}%
\end{eqnarray*}
Concerning the local Hamiltonian at the O site, it only includes the
on-site energy term which is introduced to take into account the energy
difference between the occupied orbitals of O and TM.
\begin{equation}
H_{el}^O(j) =\varepsilon _{x}n_{j,px}+\varepsilon _{y}
n_{j,py}+\varepsilon_{z}n_{j,pz}
\end{equation}

Furthermore, we consider the TM-oxygen hopping, which includes all the
allowed symmetry terms according to the Slater-Koster rules
\cite{Slater1954,Brzezicki2015} for a given bond connecting a TM to an
oxygen atom along a given symmetry direction, e.g. the $x$-axis. Here,
we allow for rotation of the octahedra around the $c$-axis assuming
that the TM$-$O$-$TM bond can form an angle $\theta$ as depicted in Fig.
\ref{fig:1}. The case with $\theta=0$ corresponds to the tetragonal
undistorted bond, while a non-vanishing value of $\theta$ arises when
the TM$-$O$_6$ octahedra are rotated at the corresponding angle around
the $c$-axis.

For the TM$-$O hopping term, we assume that for a generic bond
connecting the TM to the O atoms along the $x$-direction, the $d-p$
hybridization includes all the allowed terms, and thus
\begin{eqnarray*}
H_{{\bf TM}_1-{\bf O}}[x] &=& t_{d_{\alpha},p_{\beta}}\left(
d_{i,\alpha\sigma }^{\dagger }p_{i+a_{x},\beta\sigma}+H.c.\right)
\end{eqnarray*}
where the hopping $t_{d_{\alpha},p_{\beta}}$ is a function of the bond
angle $\theta$.
In particular, according to the Slater-Koster rules, one has:
\begin{eqnarray*}
t_{d_{xy},p_x}&&=
\sqrt{3}*n_x^2*n_y*V_{pd\sigma}+n_y (1-2 n_x^2)*V_{pd\pi}, \\
t_{d_{xy},p_y}&&=
\sqrt{3}*n_y^2*n_x*V_{pd\sigma}+n_x (1-2 n_y^2)*V_{pd\pi}, \\
t_{d_{xz},p_z}&&=n_x*V_{pd\sigma}, \\
t_{d_{yz},p_z}&&=n_y*V_{pd\pi},
\end{eqnarray*}
with $n_x=\cos \theta$ and $n_y=\sin \theta$.
In a similar way, one can
also express the $p-d$ hybridization amplitude for the O-TM$_2$ bond.
By symmetry correspondence, one can write down the other hybridization
terms along the other TM-O-TM bonds for the $y$-direction.

Since we are interested in the total free energy at finite temperature
as a function of the bond tilting angle, it is useful to introduce the
expression which is generally given by
\begin{equation}
F=-\frac{1}{\beta}\ln\left\{\sum_i\exp\left(-\beta E_i\right)\right\}\,,
\end{equation}
where $\beta=1/k_B T$ is the inverse temperature with $k_B$ being the
Boltzmann constant, while $E_i$ are the eigenvalues of the Hamiltonian
evaluated by exact diagonalization.

To proceed further with the analysis, we determine the whole energy
spectrum and the corresponding eigenstates for the TM$-$O$-$TM cluster.
Having in mind the Ca$_2$RuO$_4$ system, we start our analysis by
fixing the atomic electronic parameters in a regime matching with the
AF ground state of the monolayer compound
\mbox{\cite{Mizo01,Das18,Sutter17,Jain17}}.
{\color{black}{\noindent Furthermore, the total number of electrons is $N_e=14$ for the TM($d^4$)-O($2p^6$)-TM($d^4$) configuration, while it is $N_e=13$ for doped TM($d^3$)-O($2p^6$)-TM($d^4$) case.}

There, the severe flattening of the RuO$_6$ octahedron occurring below the structural transition,
corresponds to negative values for $\delta$, while, according to first
principle calculations or estimates employed to reproduce the resonant
inelastic x-ray and the magnetic properties, its amplitude is in the
range $|\delta|\in[200,300]$ meV \cite{Das18,Porter18}.

Furthermore, it is useful to point out that material-specific values
such as $\lambda=0.075$ eV, $U\in[1.5,2.5]$ eV, and $J_H\in[0.35,0.5]$
eV is taken as a reference for the current study. Similar values for
$\delta$, $U$ and $J_H$ have been used for calculations of electronic
spectra in Ca$_2$RuO$_4$ and the ratio $g=\delta/(2\lambda)$ is
typically considered to lie in the range $\in[1.5,2]$ for modeling the
spin excitations observed by a neutron, Raman, and resonant inelastic
x-ray scattering \cite{Das18,Arx20,Souliou17,Gretarsson19}. For the
hopping amplitudes, according to first principle calculations, one can
assume the $p-d$ values to be in the range of [1.0,2.0] eV
\cite{Malvestuto13}. These values are in the range of those employed to
describe the electronic and magnetic properties of ruthenate oxides
\cite{Forte16,Malvestuto13,Vee14}.

\begin{figure}[t!]
\includegraphics[width=0.44\textwidth]{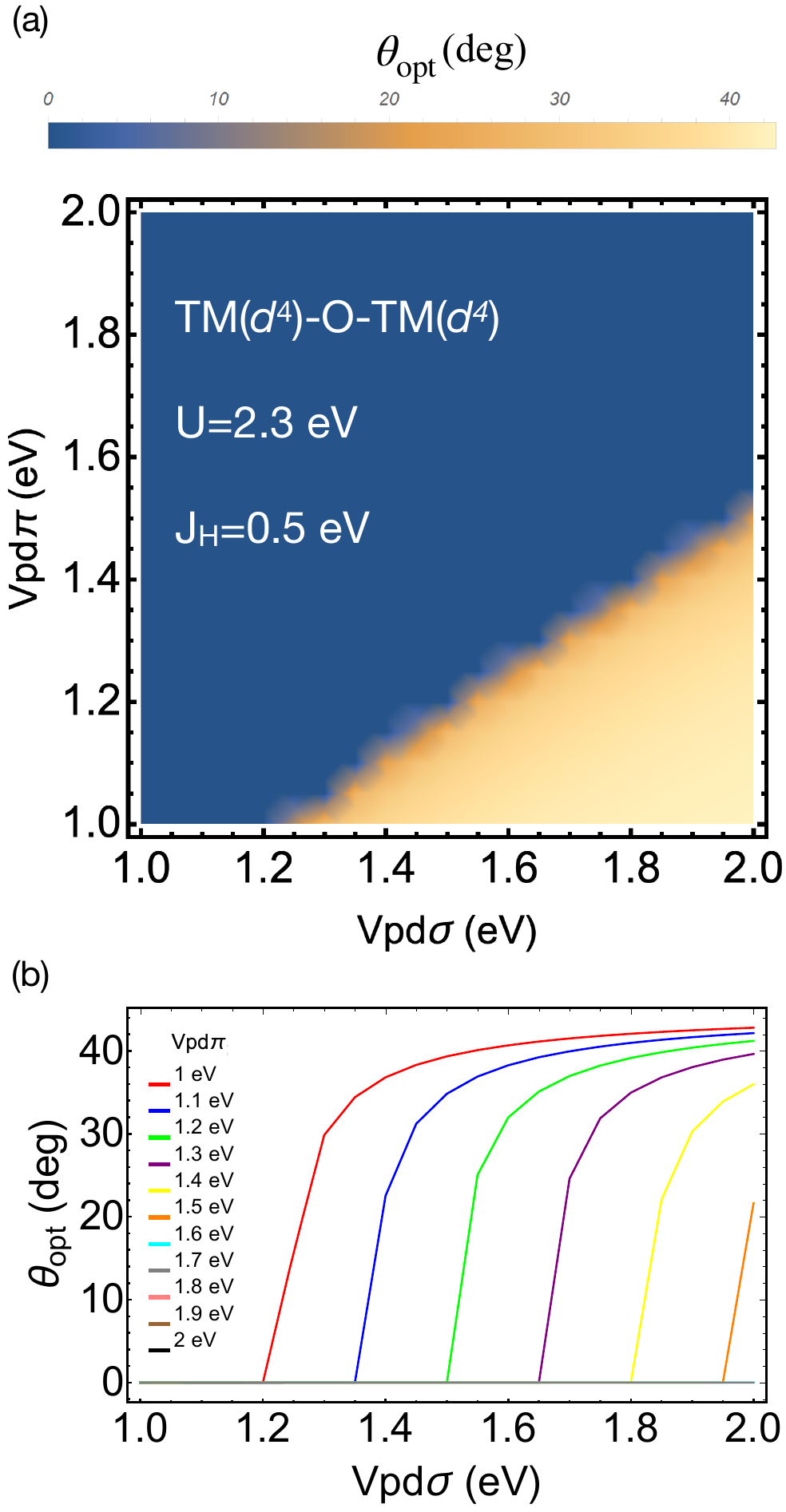}
\caption{(a) Contour map of the optimal bond angle for the
TM($d^4$)$-$O$-$TM($d^4$) configuration at zero temperature as a
function of the $p-d$ hybridization parameters
$\{V_{pd\sigma},V_{pd\pi}\}$ for
a representative value of the Coulomb strength and Hund coupling, i.e.,
$U=2.3$ eV and $J_{H}=0.5$ eV. We notice that there are two distinct
regimes of bond distortions that can be favored by the electronic
correlations. For $V_{pd\pi}$ about larger than $V_{pd\sigma}$, the
the tendency is to favor a small tilt of the bond angle which is close
to zero. Otherwise, in the remaining portion of the phase diagram the
the trend is to stabilize a large bond angle of $\pi/4$. The transition
between the two regimes is quite rapid and occurs in the proximity of
the line $V_{pd\sigma}\sim V_{pd\pi}$. A variation of the electronic
parameters do not alter significantly the phase diagram.(b) Evolution of the optimal angle as a function of $V_{pd\sigma}$ for fixed values of $V_{pd\pi}$, for the same choice of parameters as in (a).}
\label{fig:2}
\end{figure}

\subsection{Undoped TM($d^4$)$-$O$-$TM($d^4$) bond}

In Fig. \ref{fig:2} we present the distribution of the ground state
optimal bond-angle $\theta_{\rm opt}$ for the TM($d^4$)$-$O$-$TM($d^4$)
configuration at zero temperature as a function of the $p-d$ hybridization
parameters $\{V_{pd\sigma},V_{pd\pi}\}$, assuming a representative value
of the Coulomb strength and Hund's coupling, i.e., $U=2.3$ eV and
$J_H=0.5$ eV.
{\color{black}{\noindent For the oxygen orbitals we assume that the
atomic energy is given by $\epsilon_{x,y,z}=-4.5$ eV.}
This analysis allows us to determine the role of the $p-d$
covalency in setting out the distortions on the basic unit represented
by the TM$-$O$-$TM bond. There are two distinct regimes of bond
distortions that can be favored by the electronic correlations. For
$V_{pd\pi}$ about larger than $V_{pd\sigma}$, one finds a tendency to
favor a small tilt of the bond angle with an amplitude that is close
to zero. Otherwise, in the remaining portion of the phase diagram, the
trend is to stabilize a large TM$-$O$-$TM bond angle, of $\pi/4$
(then all bonds are $90^\circ$). The transition between the two
regimes is rapid and generally occurs in proximity to the line
$V_{pd\sigma} \sim V_{pd\pi}$.

\begin{figure}[t!]
\includegraphics[width=1.0\columnwidth]{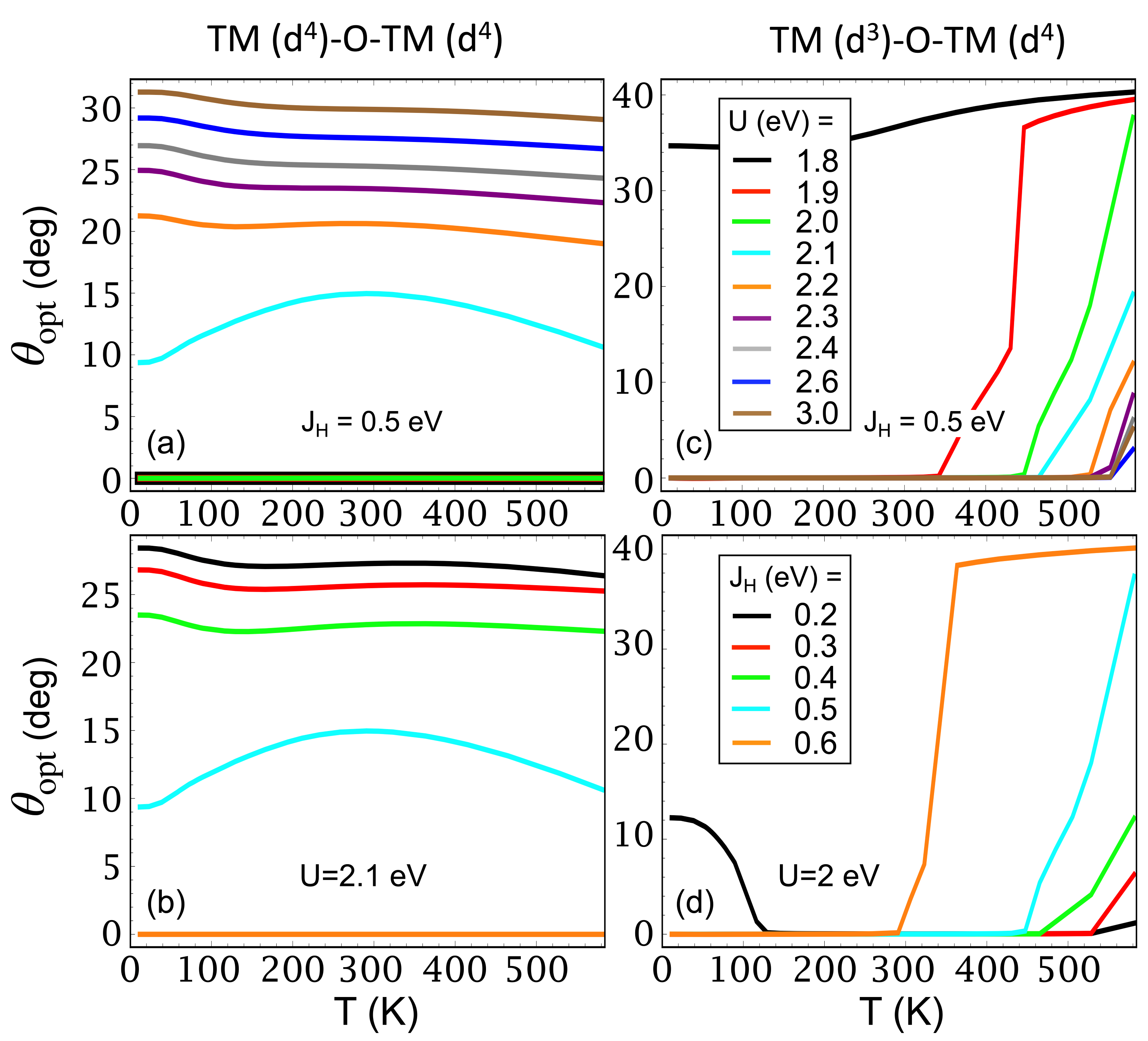}
\caption{
Temperature dependencies of the bond angles for:
(a)\&(b) TM($d^4$)$-$O$-$TM($d^4$) and
(c)\&(d) \mbox{TM($d^3$)$-$O$-$TM($d^4$)} bonds.
Curves in (a) and (c) are for fixed $J_H$ and different values of $U$
and in (b) and (d) are for fixed $U$ and different $J_H$, the values of
$U$ and $J_H$ are indicated. The other parameters are:
$V_{pd\pi}=1.3$, $V_{pd\sigma}=1.6$, $\lambda$=0.075, $\delta=0.25$,
 $\delta_{ort}=0.09$, all in eV.}
\label{fig:3}
\end{figure}

We point out that a similar trend is also obtained for the
TM($d^3$)$-$O$-$TM($d^4$) bond configuration and that a variation of
the electronic parameters (e.g. $U$, $J_H$, etc.) within the physical
regions previously defined do not alter significantly the structure
of the phase diagram. This trend can be qualitatively captured by
considering the dependence of the $d-p$ hybridization on the TM-O-TM
bond angle. For large $V_{pd\pi}$ the $d-p$ charge-transfer and the
resulting kinetic energy is optimized by $\theta\simeq 0$. In the
opposite regime with large $V_{pd\sigma}$ a significant deviation from
$\theta\simeq 0$ increases the kinetic energy due to the $d-p$ hopping
processes.

Let us then consider the thermal evolution of the optimal bond angle.
To this aim, at each temperature, we determine the minimum of the free
energy with respect to the TM-O-TM bond angle. Specifically, we
select a regime of $d-p$ hybridization for which the bond angle has a
nonzero value in the ground state, thus implying that the bond is
distorted at low temperatures. Then, we track the evolution of the angle
$\theta_{\rm opt}$ that minimizes the free energy as a function of
temperature, by varying the Coulomb interaction $U$ and the strength
of Hund's exchange.

\begin{figure}[t!]
\centering
\includegraphics[width=0.49\textwidth]{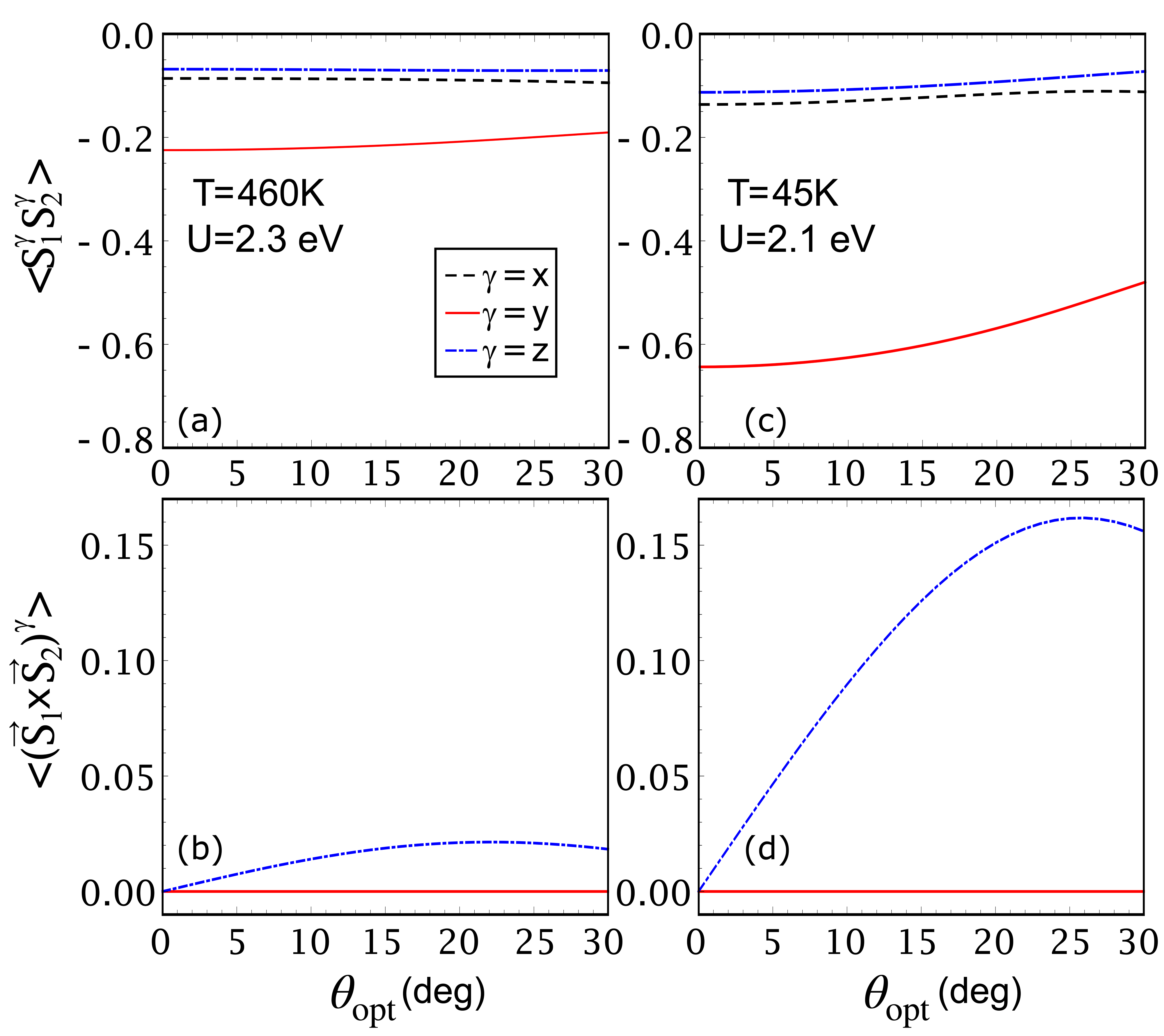}
\caption{(a), (c) Temperature dependence of the scalar product of the
spin moments at the TM sites for a \mbox{TM($d^4$)$-$O$-$TM($d^4$)}
configuration, as a function of the bond angle at $T=460$ K
and $T=45$ K for $U=2.3$ eV and $U=2.1$ eV, respectively.
(b) and (d) provide the evolution of the vector product of the spin
moments at the TM sites as a function of the bond angle at $T=460$ K
and $T=45$ K for $U=2.3$ eV and $U=2.1$ eV, respectively. }
\label{fig:4}
\end{figure}

In Figs. \ref{fig:3}(a) and \ref{fig:3}(b) we show a representative
behavior of the optimal bond angle for the TM($d^4$)-O-TM($d^4$)
configuration as a function of the temperature. As one can see by
inspection of Fig. \ref{fig:3}, the intra-orbital Coulomb interaction
$U$ is able to drive a transition from a regime where the bond angle is
insensitive to the temperature change and it optimizes the total
electronic energy by keeping the bond angle undistorted, to another
a regime of strong coupling where the high-temperature bond angle is
large and exhibits a tendency to decrease by reducing the temperature.
On the other hand, we find that the strength of Hund's exchange also
affects the critical value of the Coulomb interaction where the
transition to an undistorted bond angle occurs, such transition being
favored for higher values of $J_H$. Moreover, it affects the crossover
the temperature associated with a variation of the bond angle. Such
dependence on Hund's coupling clearly indicates that the spin and
orbital correlations are linked to the thermal changeover of the bond
angle configuration.

In order to assess the microscopic mechanisms behind the setting of the
energetically most favorable bond angle for the TM$-$O$-$TM
configuration, we focus on the spin correlations between the spin
moments at the TM sites by looking at both the scalar and vector product
between spin operators. This study aims to search for a link between the
spin configuration and the optimal bond distortion.

In Fig. \ref{fig:4} we report few representative cases for
TM($d^4$)$-$O$-$TM($d^4$) configuration to illustrate the overall trend.
The bond-angle evolution of the thermal
average of the scalar product between the spin moments at the TM sites
(i.e., the Heisenberg interaction) indicates that the amplitude is
almost temperature independent when $x$ or $z$ spin-spin correlation is
considered, whereas $y$-$y$ spin correlation tends to become of AF-type
with decreasing the temperature [see Figs. 4(a) and 4(c)], if the bond
becomes undistorted, i.e., when $\theta=0$. Moreover, looking at the
vector spin-spin correlation, we note that its out-of-plane component
gets larger than the planar one, whose value is almost vanishing. We
also notice that the amplitude of the bond angle that minimizes the free
energy does not correspond to an angle configuration that maximizes the
spin correlations for both the parallel and perpendicular spin
configuration.

\subsection{Hybrid TM($d^3$)-O-TM($d^4$) bond}

In Figs. \ref{fig:3}(c), \ref{fig:3}(d) we consider the behavior of the
hybrid TM($d^3$)-O-TM($d^4$) bond, where the electron occupation has
been changed at one TM site, by replacing $d^4$ with a $d^3$ impurity.
The temperature dependence of the optimal bond angle is completely
different if compared to the TM($d^4$)-O-TM($d^4$) bond. Indeed, we find
that the general trend is to observe a decrease in the bond angle
toward a more undistorted configuration for all the considered values
of the Coulomb interaction. Although the minimization of the free energy
yields large values of the bond angle at high temperatures it is relevant
to point out the trend of the electronic driving force in distorting the
TM-O-TM bond. Interestingly, an opposite behavior is observed when $U$
increases and $J_H$ decreases; in this case indeed the bond angle
increase when the temperature is lowered.

Focusing on the spin-spin scalar and vector product correlations between
the TM sites, we find that the evolution of the thermal average of the
scalar product upon the bond angle indicates that the amplitude is
maximal when the bond is undistorted, i.e., $\theta=0$ [see Figs.
\ref{fig:5}(c) and \ref{fig:5}(d)]. On the other hand, we note that the
the amplitude of the bond angle that minimizes the free energy at low
temperatures do correspond to an angle configuration which favors
in-plane parallel spin configurations, with a prevalent component along
the bond direction.

\begin{figure}[t!]
\centering
\includegraphics[width=0.49\textwidth]{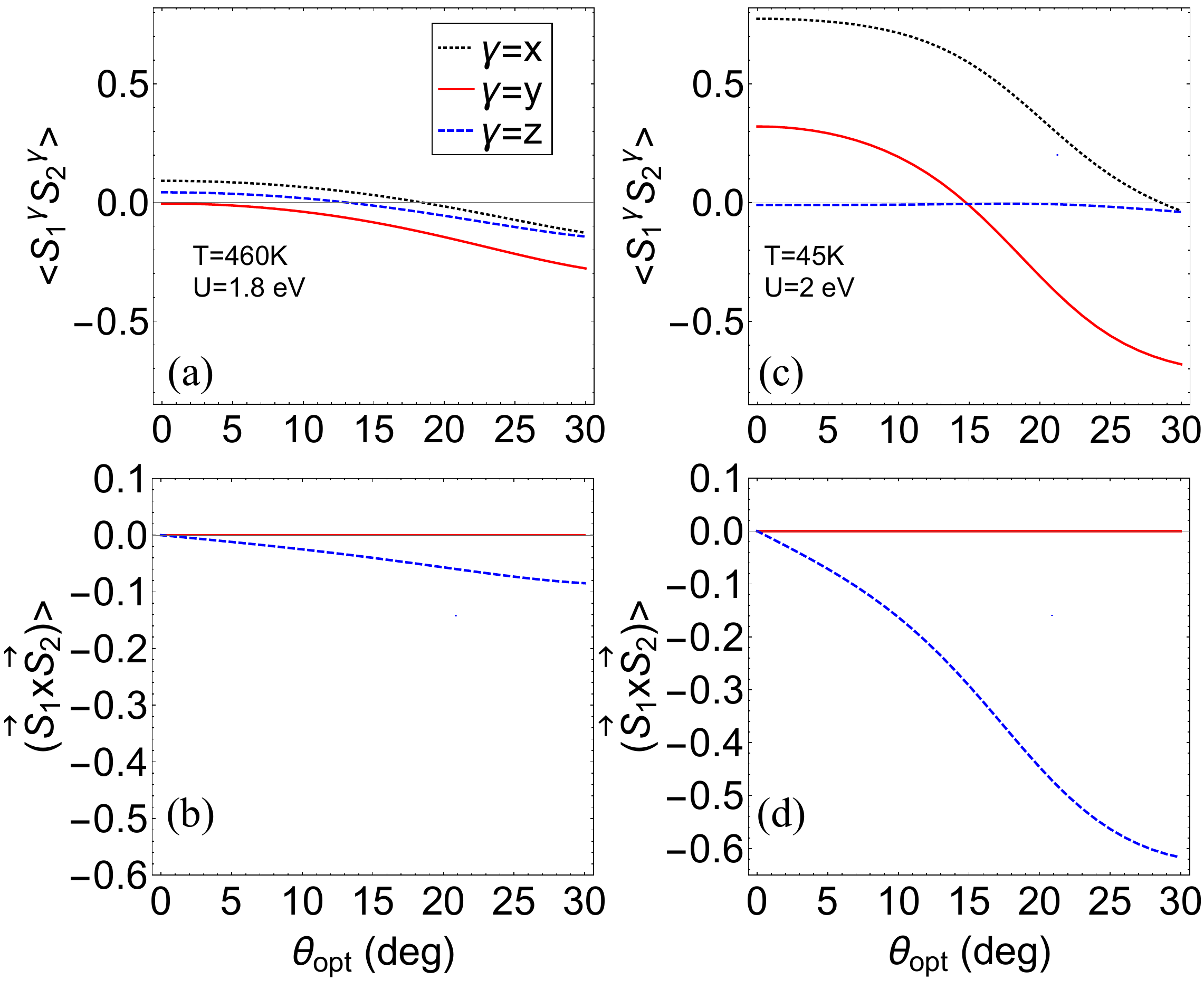}
\caption{{\color{black}(a)-(c) Temperature dependence of the expectation
value of the scalar product of the spin moments at the TM sites for a TM($d^3$)$-$O$-$TM($d^4$) configuration, as a function of the bond angle
at two representative temperatures, $T=460$ K and $T=45$ K, for $U=1.8$ eV
and $U=2.0$ eV, respectively. (b) and (d) provide the evolution of the
vector product of the spin moments at the TM sites as a function of bond
angle at $T=460$ K and $T=45$ K for $U=1.8$ eV and $U=2$ eV,
respectively.
}}
\label{fig:5}
\end{figure}

While this is expected in a single-band picture where the Heisenberg
exchange depends on the hopping amplitude, i.e. $J\sim t^2/U$, and
the charge transfer strength $t$ is typically reduced by tilting the
bond angle, in a multi-orbital scenario as that one upon examination
the competition between the various orbital channels does not allow to
have a direct prediction. On the other hand, the vector product spin
correlations (i.e., the Dzyaloshinskii–Moriya (DM) interaction) are
activated by distorting the bond as expected due to the inversion
symmetry breaking arising from the bond tilting, see
Figs. \ref{fig:5}(e) and \ref{fig:5}(f).

The general outcome is to have a maximal value of the vector product
spin correlations for a bond angle that is nonvanishing. Moreover,
as we explicitly demonstrate in Fig. \ref{fig:5}(e)\&\ref{fig:5}(f),
the maximal amplitude of the DM exchange is also temperature dependent.
These results indicate that the setting of the optimal angle in a
TM($d^3$)-O-TM($d^4$) bond is a consequence of the subtle competition
between the Heisenberg and DM interactions, which are in turn strongly
tied to the orbital degrees of freedom and to spin-orbital coupling
strength.

\subsection{Comparison between TM($d^4$)-O-TM($d^4$)
                  and TM($d^4$)-O-TM($d^3$) bonds}

The key outcomes of the TM$-$O$-$TM bond analysis can be summarized as
follows. Firstly, the study of the optimal bond angle as a function of
the temperature and Coulomb interaction indicates that the behavior for
the configuration TM($d^3$)$-$O$-$TM($d^4$) is substantially different
from the case with TM($d^4$)$-$O$-$TM($d^4$). Indeed, the
TM($d^4$)$-$O$-$TM($d^4$) is marked by an {\it increase} of the bond
angle by reducing the temperature, while for the
TM($d^3$)$-$O$-$TM($d^4$) one finds that the bond
angle gets {\it reduced} by cooling down the system from high to low
temperature. This implies that the NTE effect arising from a
modification of the bond angle, as schematically depicted in Fig.
\ref{fig:1}, is more favorable to be realized for the
TM($d^3$)$-$O$-$TM($d^4$) configuration.

A second relevant outcome of the analysis is that the optimal bond
angle for the ground state in terms of the hybridization $d-p$
parameters generally show that there are two electronic regimes
corresponding to a region of the phase space above (below) the diagonal
in the $\{V_{pd\sigma},V_{pd\pi}\}$ with small (large) bond angles,
respectively. This is a general trend and the boundary can slightly
move when varying the electronic parameters as the local Coulomb
interaction and other atomic terms as the crystal field potential.

Finally, the evolution of the magnetic state as a function of the bond
angle indicates that the configuration with minimum energy does not
correspond to a state where there are maximal spin correlations for
both the scalar and vector product between the spins at the TM sites.
This implies that the variation of the bond angle results from a
competition between the tendency to have parallel spin moments (bond
angle about zero) or perpendicular ones when the inversion symmetry
breaking is more pronounced (bond angle different from zero). In this
context turns out that spin-orbit coupling is a relevant term,
as it couples the spins to the orbital momenta and thus to the spatial
orientation of the bond angle. In the TM($d^3$)$-$O$-$TM($d^4$)
configuration, the spin-orbit coupling is inactive at the $d^3$ site,
so we argue that its suppression can account for the tendency to
observe an optimal bond angle which at low temperatures is always close
to zero.

{\color{black}{We point out that according to the relation $U^{'}=U-2J_H$, if $J_H/U>1/3$ then $J_H>U^{'}$, which in turn sets a boundary for the crossover from weak to strong Hund's coupling.
The solution of the multi-orbital Hubbard model for the single bond has been explored to investigate the changeover from weak to strong Hund's coupling regime.
Indeed, the amplitude of Hund's interaction, as demonstrated in Fig. 3 (a,b), plays a relevant role in setting out the thermal negative expansion effects.
This is evident, for instance, in the case of $d^4-O-d^4$ bond where one finds a modification of the optimal angle with a positive thermal expansion effect (i.e. increase of the bond angle at low temperature), that turns into a negative thermal one when the amplitude of Hund's coupling grows. Along this line, an enhancement of the NTE effects, with a larger negative variation of the optimal angle and a shift of its onset towards higher temperatures are also observed for the doped $d^3-O-d^4$ bond configuration (Fig. 3 (c,d)).
}

The presented analysis is applicable to all oxide materials.
However, since there is a clearcut experimental evidence of anomalous
negative volume effects in the insulating phase of the Ca$_2$RuO$_4$
compound and derived doped materials belong to the same family, we
focus on an electronic regime that is relevant for ruthenium oxides.
There, the Ru atom is in a $d^4$ configuration and one allows for
atomic substitutions that are isovalent and can lead to a change into
$d^3$ or $d^2$ electronic state at the TM site. This variation can be,
for instance, induced by replacing the Ru with Mn or Cr ions for
achieving a $d^3$ or $d^2$ configuration, respectively.

\section{Plaquette consisting of $d^{4}$ ions: low-energy spin-orbital description}
\label{sec:44}

In order to account for the electronic behavior in two dimensions, a
minimal requirement is to simulate bonds in two inequivalent directions.
To this aim, we choose to study a cluster based on a $2\times 2$
plaquette. This choice is motivated by the fact that it is highly
computationally demanding to deal with the full d-p multiband Hubbard
model as the size of the Hilbert space rapidly grows. Therefore, we
resort to an approximate treatment valid in the insulating regime with
frozen charge degrees of freedom due to a large $U$ coupling leading
to an effective spin-orbital exchange. The spin-orbital Hamiltonian for
the $d^4$ configuration with three $t_{2g}$ orbitals has been already
derived \cite{Pen97,Dag10} for a basic symmetric configuration.

Here, instead, we need two new microscopic ingredients:
(i) a non-trivial TM$-$O$-$TM bond angle, and
(ii) the oxygen degrees of freedom in the virtual excited states.
Note that (i) can be implemented without (ii) if we neglect the oxygen
degrees of freedom so the effect of bond distortion can, to some
extent, be captured by a simplified model. Nevertheless, we use the
full model with oxygens because of the lack of the oxygen degrees of
freedom cannot properly describe a lattice configuration with a
variable bond angle. Indeed, we have found that the simplified model
with projecting out oxygen degrees of freedom never favors a non-zero
bond angle in the ground state which does not match the results of
the $d-p$ multi-band Hubbard model.

The detailed derivation of the TM$-$O$-$TM spin-orbital exchange up to
the fourth order is given in Appendix A. The resulting Hamiltonian
couples two spins $S=1$ and orbitals $L=1$ along a bond. The bond
Hamiltonian depends on the direction of the bond in the lattice and the
bond angle. We adopt an idealized $2\times 2$ plaquette with fixed
\mbox{TM$-$O} distances and the distances between TM ions being reduced
by cooperative distortions. The schematic view of the plaquette and the
convention chosen for the site indices and for the angles indicating the
rotations of RuO$_6$ octahedra are given in Fig.~\ref{fig:6}. Therefore,
the Hamiltonian for the $2\times 2$ cluster of four equivalent $d^4$
ions can be written as,
\begin{eqnarray}
{\cal H}&\!=&{\cal H}_{1,2}\left(\theta\right)\!
+\!{\cal H}_{2,4}\left(-\tfrac{\pi}{2}\!+\!\theta\right)\!
+\!{\cal H}_{4,3}\left(-\theta\right)\!
+\!{\cal H}_{3,1}\left(\tfrac{\pi}{2}\!-\!\theta\right)\! \nonumber\\
+&\!&\!\!\!\!\!\lambda\!\sum_{i=1}^4 \vec{L}_i\!\cdot\!\vec{S}_i \!
- \sum_{i=1}^4\! \left[(\delta-\delta_{ort}) a^{\dagger}_i a_i \!
+\!(\delta+\delta_{ort}) b^{\dagger}_i b_i  \right]\!.
\label{Hamd4}
\end{eqnarray}
where $a^{\dagger}_i$, $b^{\dagger}_i$ and $c^{\dagger}_i$ are
the hard-core boson operators that create a double occupation
of the $yz$, $zx$ and $xy$ orbital, respectively.

\begin{figure}[t!]
\includegraphics[width=1\columnwidth]{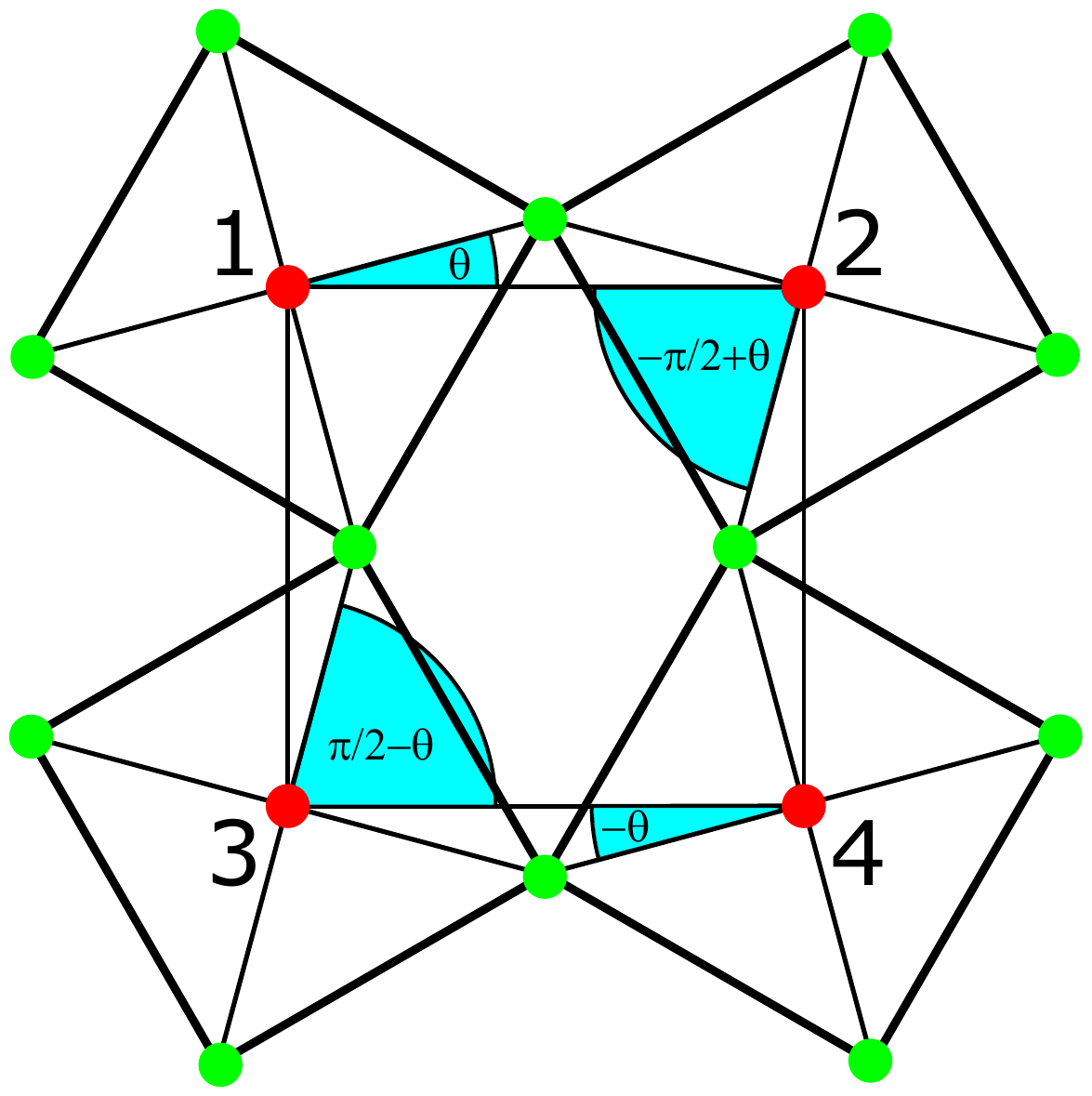}
\caption{Schematic view of a plaquette of four metallic sites (red dots)
with surrounding oxygens (green dots) with cooperative rotational
distortions by angle $\pm\theta$. The bond angles are explicitly
highlighted in cyan.
\label{fig:6}}
\end{figure}

We study the ground state and finite-temperature properties of the
Hamiltonian (\ref{Hamd4}), namely the behavior of the bond angle
$\theta$ that minimizes the free energy of the system as given by
\begin{equation}
F(\theta,T)=-\frac{1}{\beta}\ln\left\{
\sum_n\exp\left(-\beta E_n(\theta)\right)\right\},
\label{eq:ffree}
\end{equation}
where the sum runs over the eigenvalues $\{E_n(\theta)\}$ of ${\cal H}$.
We also investigate the behavior of the bond angle in presence of a
magnetic order with ferromagnetic alignment along one direction and AF
in the other, thus realizing a so-called stripe $C$-AF pattern. This
configuration is stabilized by adding an effective Zeeman field that
breaks the symmetry and aligns the spin $S$ and orbital $L$ angular
moments along the chosen orientation $z$:
\begin{equation}
{\cal H}_{\rm mag}=h (J^z_1+J^z_2-J^z_3-J^z_4),
\end{equation}
with $\vec{J}_i = \vec{S}_i + \vec{L}_i$.

In Fig. \ref{fig:7}(a) we show a phase diagram of a plaquette as a
function of the hybridization amplitudes $\pi$ and $\sigma$, per analogy
to the one obtained for a single bond shown in Fig. \ref{fig:2}. Here,
at first glance, we see only two phases: the one with zero bond angle
(straight bonds) when $\pi$ hybridization is dominating, and the one
with $\pi/4$ bond angle when $\sigma$ hybridization dominates. However,
looking closer one notices a very narrow intermediate phase along the
phase boundary when the bond angle takes intermediate values. This is
indicated by the behavior of a representative free energy curve found
in this phase, see Fig. \ref{fig:7}(b), having a minimum for an
intermediate value of the optimal bond angle $\theta_{\rm opt}$. One
can then immediately track the behavior when going from one phase to
the other.
{\color{black}{We point out that this peculiar behavior of the optimal angle, with only a small window of parameters for the $d-p$ hybridization amplitude showing a rapid variation from small to large angles, can be regularized by including the effects of the lattice potential. This effect is demonstrated in the Appendix E. It is shown indeed that, depending on the electron-lattice coupling amplitude, one can enlarge the range of electronic parameters for the $d-p$ hopping amplitudes where a smooth variation from large to small angle takes place.}}

\begin{figure}[t!]
\includegraphics[width=1\columnwidth]{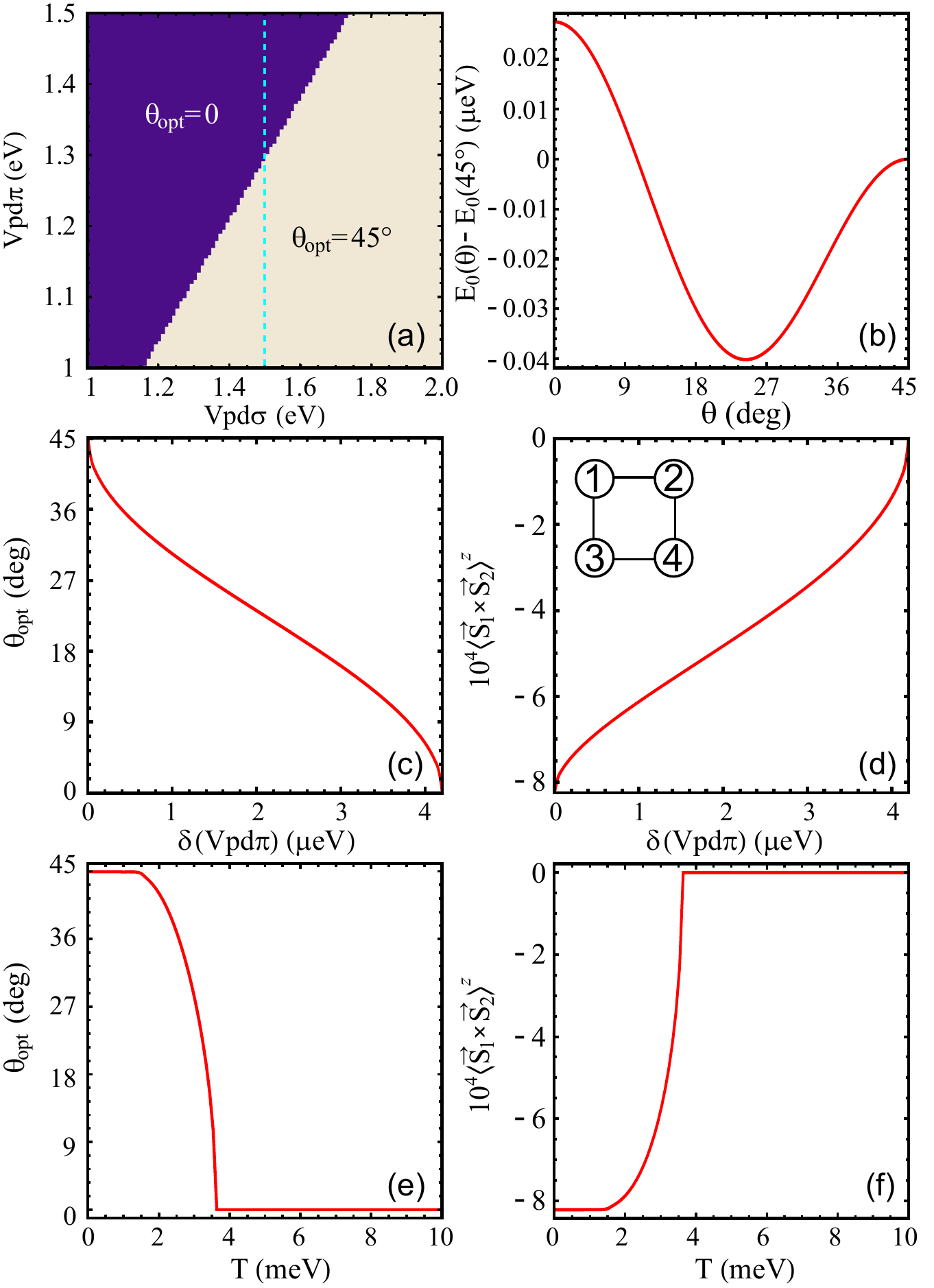}
\caption{
Results for the undoped $d^4$ plaquette [inset in (d)]:
(a)~phase diagram as function of $\pi$ and $\sigma$ hybridization
amplitudes
(b) typical curve of ground state energy versus bond angle $\theta$ in
the intermediate phase;
(c,d) typical curve of the optimal bond angle and the ground state
average of $\langle\vec{S}_1\times\vec{S}_2\rangle^z$ versus
$\delta(V_{pd\pi})\equiv V_{pd\pi}-V_{0}$ for $V_{pd\sigma}=1.5$ eV in
the intermediate phase marked by a dashed line
in plot (a);
(e\&f) thermal dependencies of the optimal bond angle and average
$(\vec{S}_1\times\vec{S}_2)^z$ in the intermediate phase for
$V_{pd\sigma}=1.5$ and $V_{pd\pi}=1.2973$ eV.
The other parameters are: $U=8.0$, $J_H=0.5$, $\epsilon_p=-4.5$,
$\delta=0.35$, $\delta_{ort}=0.09$, and $\lambda = 0.075$, all in eV.
\label{fig:7}}
\end{figure}

In Fig. \ref{fig:7}(c) we show the curve of the optimal bond versus
$pd-\pi$ hybridization when crossing the phase boundary. We observe that it
interpolates smoothly between $\pi/4$ and $0$, leading to a
second-order type transition between the two phases. Interestingly, when
searching for an observable that would follow the behavior of the
optimal angle we find that it is optimally reproduced by the
cross-product of the spins along a given nearest-neighbor bond. From the
symmetry property of the system, it follows that only the $z$ component
of such a cross-product takes non-vanishing values. Hence, in Fig.
\ref{fig:7}(d) we show the average of $(\vec{S}_1\times \vec{S}_2)^z$
in the ground state (note that here all the bonds of the plaquette are
equivalent). We see that the profile of the curve resembles very
closely the shape of $\theta_{\rm opt}$. The non-vanishing average of
the cross-product of spins means that the system tends to stabilize a
non-collinear magnetic order. Therefore, we can conclude that in the
examined plaquette the change of the optimal bond angle is closely
related to the tendency of the system to develop a non-collinear
spin texture.

\begin{figure}[t!]
\includegraphics[width=1\columnwidth]{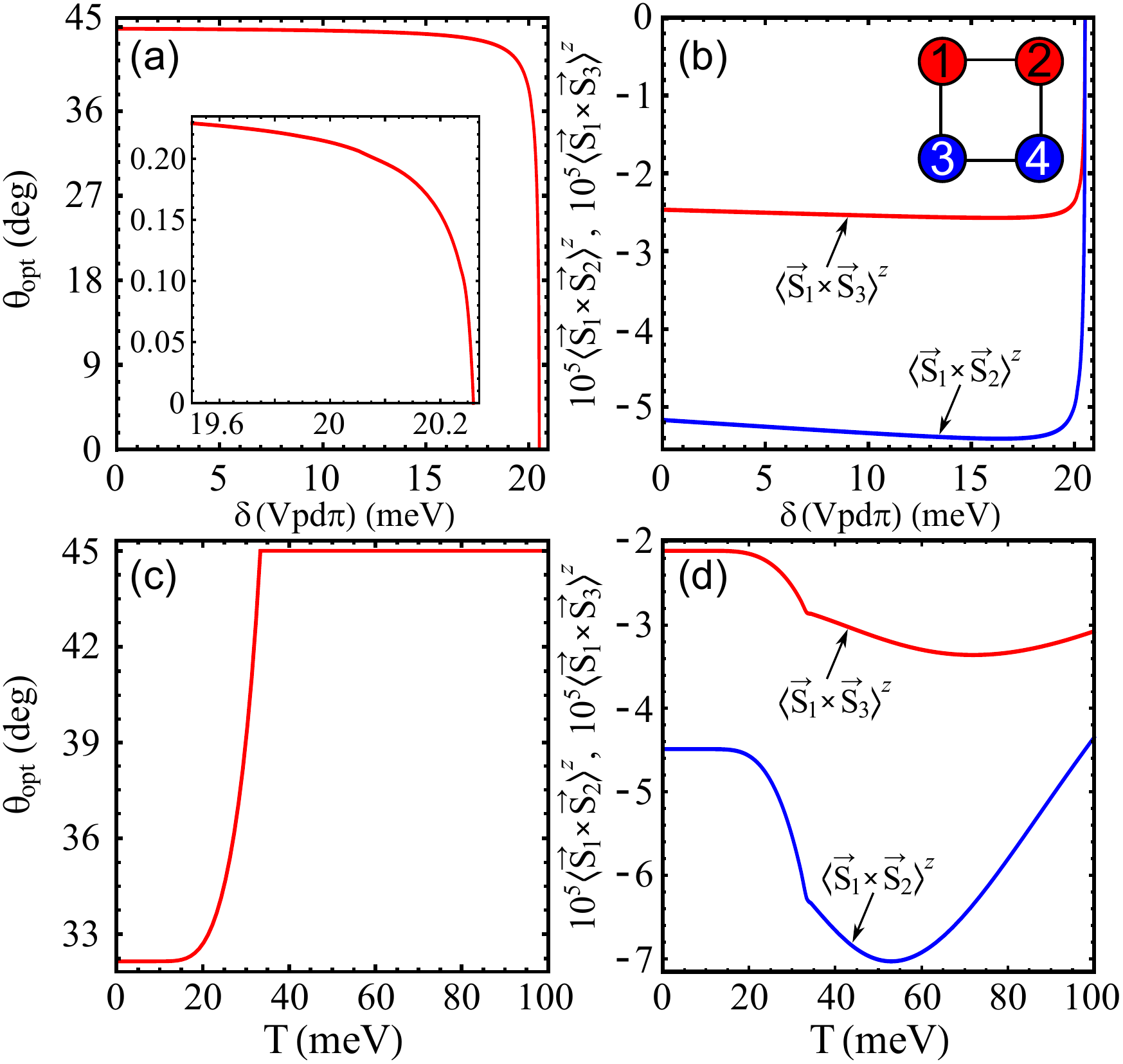}
\caption{
Results for pure-$d^4$ plaquette with $C$-AF magnetic order (see the
inset):
(a,b) typical curves of the optimal bond angle and the ground state
averages of $\langle\vec{S}_1\times\vec{S}_2\rangle^z$ and
$\langle\vec{S}_1\times\vec{S}_3\rangle^z$ versus
$\delta(V_{pd\pi})=V_{pd\pi}-V_{0}$ for $V_{pd\sigma}=1.5$ eV in the
intermediate phase between $\theta=0$ and $\theta=\pi/4$ (see Fig.
\ref{fig:6});
(c,d)~thermal dependencies of the optimal bond angle and average
$\langle\vec{S}_1\times\vec{S}_2\rangle^z$ and
$\langle\vec{S}_1\times\vec{S}_3\rangle^z$ in the intermediate phase for
$V_{pd\sigma}=1.5$ and $V_{pd\pi}=1.3034$ eV. The other parameters are:
$U=8.0$, $J_H=0.5$, $\epsilon_p=-4.5$, $\delta=0.35$,
$\delta_{ort}=0.09$, $\lambda=0.075$ and $h=0.2$, all in eV.
\label{fig:8}}
\end{figure}

The presence of an intermediate phase suggests a frustrated configuration
of the plaquette in a well-defined parameter area. Thus, we study the
thermal behavior of the system for the intermediate phase with the goal
to achieve a configuration with a NTE character. Here, the manifestation of this effect can be
tracked by identifying the configurations with a positive derivative of
the optimal angle with respect to temperature, i.e., the system should
become more distorted (bond angle different from zero) when the
temperature grows. In Fig. \ref{fig:7}(e) we show the behavior of the
optimal angle as a function of temperature in the region of the parameter that
is in close proximity to the intermediate phase. We see that initially
$\theta_{\rm opt}$ stays constant when the temperature grows up to about
$T= 1.5$ meV but then rapidly drops to zero at around $T=4.0$ meV and
stays zero until reaching a high-temperature regime. This means that no
NTE effect can be obtained in this regime of parameters. Moreover, the
behavior of $\theta_{\rm opt}$ is non-analytical at the two transition
temperatures and this behavior is again well reproduced by the chiral spin correlations shown in Fig. \ref{fig:7}(f). This outcome indicates that the thermal behavior of the system is also related to the
non-collinearity of the spin texture.

Finally, to trigger the NTE effect we have considered the differences
between the plaquette and the former single bond analysis where the
NTE effect was indeed obtained. We argue that the main difference
between the two systems is the underlying symmetry. The plaquette obeys
a fourfold in-plane rotation symmetry with which both main phases are
compatible. This suggests that the NTE effect could be obtained by
lowering the symmetry of the plaquette. To achieve this we have
employed the striped \mbox{$C$-AF} magnetic order introduced at the
beginning of this section.

The results at zero and finite temperatures are shown in Fig.
\ref{fig:8}---note that we do not show the phase diagram again because
it changes only slightly. The behavior of the optimal angle when
passing through the intermediate phase is shown in Fig. \ref{fig:8}(a).
Note that the angle still interpolates smoothly between $0$ and $\pi/4$
but the drop of the function is much more rapid than before. Again,
this behavior is well reproduced by the cross-product of the spins, see
Fig. \ref{fig:8}(b). Looking at the thermal behavior of
$\theta_{\rm opt}$ in the intermediate phase, see Fig. \ref{fig:8}(c),
we see a sharp increase of the angle starting roughly at $20$ meV and at
around $34$ meV. Therefore we get the NTE effect within this
temperature range. Interestingly the $\theta_{\rm opt}$ seems to be
non-analytical only at the larger temperature. Finally, we observe
again in Fig. \ref{fig:8}(d) that the spin cross-product follows the
optimal angle, including the non-analytical point.

{\color{black}{The qualitative outcome is not affected by the increase of the $J_H/U$ ratio, as inferred from the results reported in Appendix C where we have analyzed how the thermal profile of the bond angle is affected by a change of the Hund's coupling and spin-orbit coupling.
Moreover, as shown in Appendix D where we plotted the results for the low energy spin-orbital model for the case of a single bond, we find that the qualitative trend is similar to that obtained by solving the full multi-orbital model in the regime of strong $U$ coupling.}

\section{Plaquette with $d^3$ impurity}
\label{sec:34}

Now, we turn to the analysis of the bond-angle evolution with increasing
temperature causing changes in the thermal average spin correlations
developing on a TM($d^3$)$-$O$-$TM($d^4$) bond. Similarly to what has
been done in the previous section, we have derived a superexchange
Hamiltonian of the plaquette with a single $d^3$ dopant to be able to
solve the problem in the limit of large $U$. A single dopant in the $d^4$
plaquette results in two $d^4-d^4$ bonds between the host sites and
two-hybrid $d^3-d^4$ bonds between impurity and host atoms. Therefore,
the hybrid bonds need to be derived including the effect of the oxygen
degrees of freedom and bond distortions, as in the pure $d^4$ case. The full derivation of the effective model is reported in Appendix B. Note that the case with
no distortion and no oxygens was already addressed in Ref. \cite{Brz15}.

\begin{figure}[t!]
\includegraphics[width=1\columnwidth]{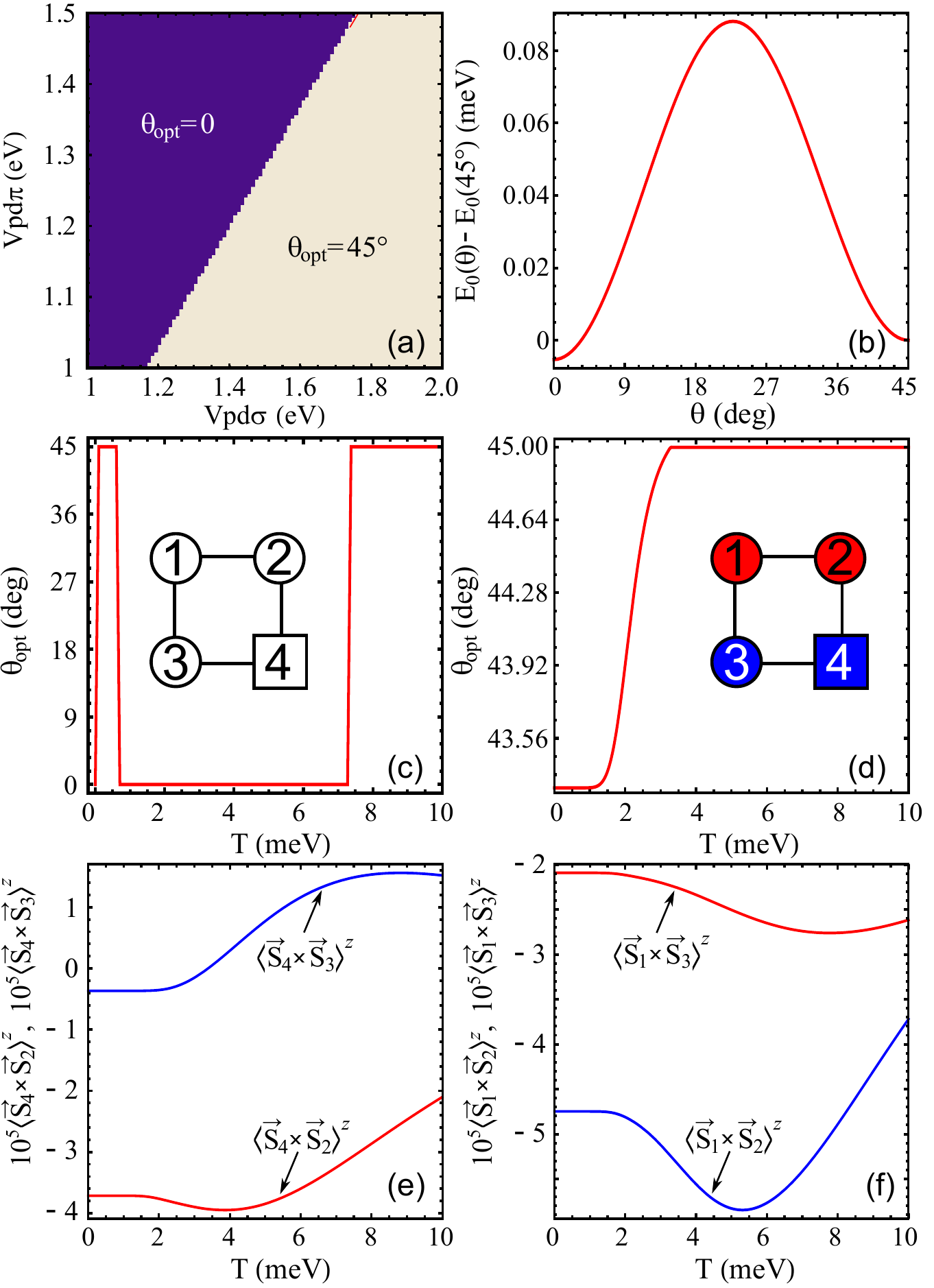}
\caption{
Results for the $d^3$-doped $d^4$ plaquette (insets, dopant at $i=4$):
(a)~phase diagram as function of $\pi$ and $\sigma$ hybridization
amplitudes,
(b) typical curve of ground-state energy versus bond angle $\theta$
close to the phase boundary,
(c) optimal angle versus temperature for $V_{pd\sigma}=1.5$ and
$V_{pd\pi}=1.2848$ eV, no magnetic order imposed,
(d) optimal angle versus temperature for $V_{pd\sigma}=1.5$ and
$V_{pd\pi}=1.2745$ eV, $C$-AF magnetic order is imposed as shown by
red (blue) color of the nearest neighbor bonds,
(e\&f) averages of $(\vec{S}_i\times \vec{S}_j)^z$ along the NN bonds
versus temperature for the same parameters as in (d).
The other parameters are: $U=\tilde{U}=8.0$, $J_H=\tilde{J}_H=0.5$,
$I_e = 1.0$, $\epsilon_p=-4.5$, $\delta = 0.35$, $\delta_{ort}=0.09$,
and $\lambda = 0.075$, all in eV.
\label{fig:9}}
\end{figure}

It is worth mentioning here that while for pure host bonds, having large
$U$ alone is enough to stabilize the low-energy spin-orbital model, in
the case of the hybrid bonds it is necessary to add a mismatch between
the atomic levels of the $d^3$ atom with respect to the $d^4$ one in
order to secure different valences of these atoms. As a result, one gets
an exchange model between spin $S=1$ and orbital $L=1$ at one site and
the impurity spin \mbox{$S=3/2$} with no orbital degree of freedom at
the impurity. Taking the same angle convention as that shown in Fig.
\ref{fig:6}, the Hamiltonian with a $d^3$ impurity at site $i=4$ reads
as follows [cf. Eq. \eqref{Hamd4}],
\begin{eqnarray}
{\cal H}&\!=\!&{\cal H}_{1,2}\left(\theta\right)\!
+\!{\cal H}_{4,2}^{\rm imp}\left(\tfrac{\pi}{2}\!+\!\theta\right)\!
+\!{\cal H}_{4,3}^{\rm imp}\left(-\theta\right)\!
+\!{\cal H}_{3,1}\left(\tfrac{\pi}{2}\!-\!\theta\right)\! \nonumber\\
+&&\!\!\!\!\! \lambda\sum_{i=1}^3 \vec{L}_i\!\cdot\!\vec{S}_i \!
-\! \sum_{i=1}^3\! \left[(\delta-\delta_{ort}) a^{\dagger}_i a_i \!
+\!(\delta+\delta_{ort}) b^{\dagger}_i b_i  \right]\!.
\label{Hamd34}
\end{eqnarray}
%

\begin{figure}[t!]
\includegraphics[width=1\columnwidth]{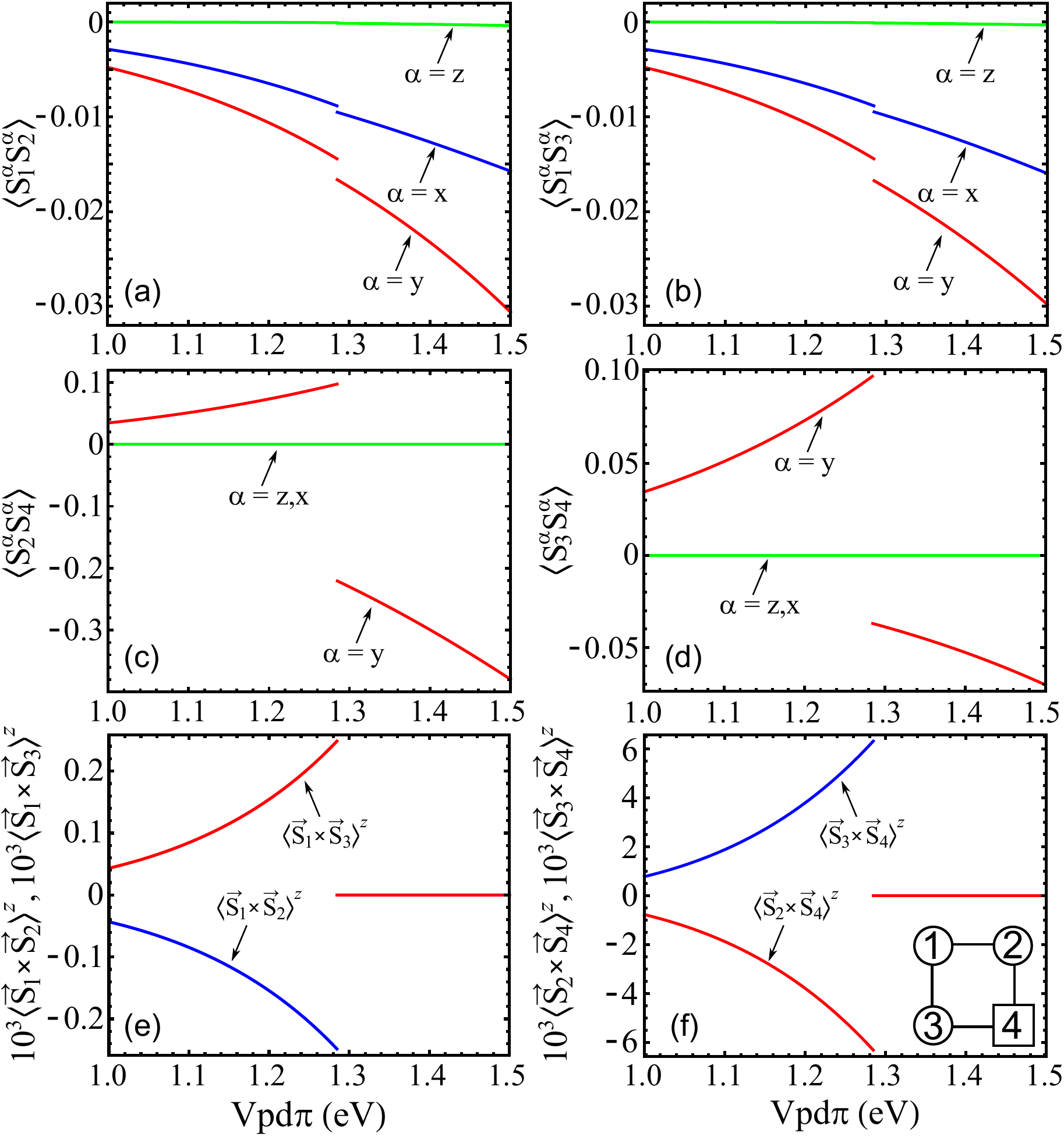}
\caption{
Spin-spin correlations for the $d^3$-doped $d^4$ plaquette (insets,
dopant at $i=4$) as functions of $\pi$ hybridization amplitude across
the phase transition between $\theta=0$ and $\theta =\pi/4$ phases at
zero temperature.
(a-d) Correlations $\langle S_i^{\alpha}S_j^{\alpha}\rangle$ for the
nearest neighbors (NN).
(e-f) Correlations $\langle\vec{S}_i\times\vec{S}_j\rangle^z$ along the
NN bonds. The other parameters are: $U=\tilde{U}=8.0$,
$J_H=\tilde{J}_H=0.5$, $I_e=1.0$, $\epsilon_p=-4.5$, $\delta=0.35$,
$\delta_{ort}=0.09$, $\lambda=0.075$ and $V_{pd\pi}=1.5$, all in eV.
\label{fig:10}}
\end{figure}

Here, we follow the same logic as in the previous section without $d^3$ impurity.
The results are shown in Fig. \ref{fig:9}. As in the previous case, the
phase diagram of Fig. \ref{fig:9}(a) exhibits two phases, one with zero
bond angle and the other with an angle $\pi/4$. Their positions are
similar as in the pure $d^4$ case, however, in contrast to that
configuration, we find no phase with an intermediate angle phase between
these two. This follows from the behavior of the free energy, see Fig.
\ref{fig:9}(b), which exhibits two local minima at $\theta =0$ and $\theta=\pi/4$. When one passes through the phase boundary the positions
of the minima do not change but the values of the minima do. As a result,
we get a discontinuous, first-order transition between the two phases.

In this context, the impact of having a finite temperature is quite
peculiar. If we place ourselves close to the phase boundary we find that
the system jumps between two optimal angles $\theta=0$ and
$\theta=\pi/4$ as the temperature grows, see Fig. \ref{fig:9}(c). We
argue that in the infinite system there will be phase separation at the
temperatures when two phases are degenerate which will make the jumps
smoother, leading to the temperature intervals hosting the NTE effect
with NTE.

The situation becomes more evident in the $C$-AF phase. The transition
at zero temperature remains first order but the optimal angle first
drops slightly from the $\pi/4$ value before jumping to zero when
crossing the phase boundary. If we now place ourselves in the region
where the angle is decreased and look at the thermal behavior, see Fig.
\ref{fig:9}(d), we get a NTE effect in the temperature range roughly
between 1.5 and $3.2$ meV. The $\theta_{\rm opt}$ curve shows a
non-analytic behavior at the latter point. Unlike in the pure $d^4$
case, looking at the spin-spin cross products, see Figs.
\ref{fig:9}(e-f), we do not observe that the curves follow
$\theta_{\rm opt}$ very closely, however in the NTE interval they
exhibit high gradients. Note that due to the presence of both the
impurity atom and the $C$-AF order, all the bonds are now inequivalent.
The low and high-angle phases shown in Fig. \ref{fig:9} differ in a
rather subtle way.

Looking at the spin-spin correlations in Fig. \ref{fig:10} as a function
of the $\pi$ hybridization, we can clearly see the discontinuities
across the transition. The
$\langle S^{\alpha}_i\!\cdot\!S^{\alpha}_j\rangle$ correlations along
the host bonds exhibit only small jumps and do not change much whereas
at the impurity bonds the changes are more notable, see Figs.
\ref{fig:10}(a-d). Specifically, the behavior of the
$\langle S^y_iS^y_j\rangle$ correlator around the impurity is
interesting---it changes signs across the transition. This change in sign
is also visible in the non-collinear
$\langle\vec{S}_i\times\vec{S}_j\rangle^z$ correlator which grows in
absolute values as we increase $V_{pd\pi}$ towards the transition just
to drop to zero and stay at zero in the entire zero-angle phase.

{\color{black}{As discussed in Sec. IV, we point out that for the doped configuration the thermal profile of the bond angle is not qualitatively affected by a change of the Hund's coupling and spin-orbit coupling (see Appendix C). Moreover, the outcomes of the single effective $d^3-d^4$ bond within the spin-orbital model are qualitatively consistent with those of the full multi-orbital Hubbard model (as reported in Appendix D).}

\section{Summary and conclusions}
\label{sec:summa}

{\color{black}{

The phenomenon of NTE effect deals with the increase of lattice distances and, in turn, of the volume of the unit cell, if the material is thermally cooled.
Here, we studied the conditions for obtaining negative thermal effects in materials characterized by strong electron-electron correlations involving spin degrees of freedom and multiple orbitals.
In this context, when considering for instance perovskite oxides with transition metal elements inside the oxygen octahedral cages, the change of the unit cell size is linked to the octahedral rotations and thus to the amplitude of the TM-O-TM bond angle. An undistorted (distorted) TM-O-TM bond with zero (non-vanishing) bond angle corresponds to  large (small) volume of the unit cell, respectively. Hence, the TM-O-TM bond angle is a key parameter in the phenomenon of NTE. The TM-O-TM bond angle, on the other hand, sets out the electrons connectivity within the crystal lattice through the hybridization of the orbitals at the transition metal and oxygen atoms. Hence, the amplitude of the electronic kinetic energy is determined by the TM-O-TM bond angle. For the case of an electron correlated insulating phase with multi-orbital configurations, the Coulomb interaction prevents the charge motion and thus the kinetic energy is substantially given by virtual charge transfer processes and by the resulting spin-orbital exchanges.
Then, while for a single orbital model one would expect that the kinetic energy is optimal by having an undistorted bond, the resulting configuration in the case of multi-orbital electronic state cannot be a priori predicted and requires the knowledge of the orbital dependent magnetic correlations.
This aspect underlines the complexity of the fundamental problem we have been facing in this paper: to establish how the spin-orbital correlations determine the TM-O-TM distortions and single out the conditions for achieving NTE effects.
\\
On this basis, it is evident that one needs to evaluate the dependence of the spin-orbital exchange on the bond angle. For this reason, we employed a theoretical approach that explicitly includes the bond angle in the amplitude of the $d-p$
hybridization processes. In this way, we have accounted for the modification of the free energy as a function of the TM-O-TM bond configuration and assess the role of the electron correlations in a multi-orbital electron system.
\\
From our study we have been able to identify few mechanisms which are crucial for achieving NTE effects.
First, the anisotropy of the orbital correlations, as pointed out above, can play an important role. In particular, the orbital anisotropy or preferential orbital occupation, which is driven by the electron correlations, leads to a ground state that is prone to exhibit NTE effects. The occurrence of these anisotropic orbital correlations drives the bond direction to get less distorted bond. This is what one can find at low temperatures where the spin-orbital correlations are stronger. On the other hand, thermal fluctuations tend to weaken the spin-orbital correlations, a condition which favors more distorted bonds. The thermal disorder also manifests in the spin channel with the formation of non-colinear magnetic correlations that energetically support a distorted bond.
If such conditions can be realized, then the optimal bond angle will decrease by increasing temperature and thus we will have resulting NTE effects.
\\
A second key element that emerges from our analysis is related to the presence of low point group symmetry conditions. In the regime of strong Coulomb interaction, we find that the breaking of the $C_4$ rotation symmetry can turn a positive thermal expansion effect into a NTE behavior when tracking the TM-O-TM bond angle. In our study, this effect has been studied by considering a magnetic pattern that breaks the $C_4$ rotation symmetry. However, we expect that similar outcomes for the NTE can be also obtained by considering orbital or electronic patterns that reduce the point group symmetry of the system. In this respect, a significant modification of the NTE effects can be foreseen in thin films under strains or by applying electric field which would naturally provide microscopic conditions with low point group symmetry.
\\
Another relevant aspect concerns the role played by the strength of Hund's coupling to induce the occurrence of NTE effects.
The solution of the multi-orbital Hubbard model for the single bond allows us to explore a regime of strong Hund's interaction, for either weak or strong intra-orbital Hubbard $U$, since it takes into account at the same level the charge, spin and orbital degrees of freedom.
Indeed, the amplitude of Hund's coupling can play an important role in setting out the thermal negative expansion effects.
This is evident, for instance, in the case of $d^4-O-d^4$ bond where one finds a changeover of the optimal angle with a positive thermal effect profile that turns into a negative thermal one by increasing the amplitude of Hund's coupling. Additionally, we also find an enhancement of the NTE effects, with a larger negative variation of the optimal angle, and a shift of its onset towards higher temperatures, for the doped $d^3-O-d^4$ bond configuration.
Since this type of changeover from positive to NTE effects is not observed when considering the large $U$ limit, as evaluated within the low-energy model employed for the simulation of the square cluster, we argue that the Hund's coupling in the intermediate regime of $U$ interaction is relevant for the occurrence of negative thermal effects in multi-obital materials.
Although our model analysis has not been expanded to investigate a Hund's exchange driven correlated metal \cite{Medici11}, on the basis of the results of the single bond for the multi-orbital Hubbard model, we argue that the effect of large Hund's coupling can favor the occurrence of NTE. In this context, we are confident that our analysis may further stimulate the search for NTE effects in doped Hund's metal exhibiting strong correlations, as for instance in iron based materials \cite{Zhou18,Niedziela21}, or in the proximity to a metal-to-insulator orbital selective transition.
\\
Concerning real materials where our study could be applicable, we recall that the analysis has been substantially motivated by the NTE observed in the Ca$_2$RuO$_4$ compound. In this context, our analysis is able to account for the occurrence of NTE effects in the Ca$_2$RuO$_4$ family as due to the transition metal substitution. The analysis specifically refers to the substitution of Ru ($d^4$) with Mn ($d^3$). We find that both in the intermediate and in the strong coupling regimes we can achieve NTE effects close to the impurities. As mentioned above, since the impurity can act as an orbital polarizer it can sustain less bond distortions at low temperatures.
\\
Finally, we mention that the inclusion of a phenomenological electron-lattice potential, as discussed in Appendix E, allows to regularize the maximal bond distortions that are obtained within a purely electronic description and to make smoother the transitions from large to small bond angles across the phase space spanned by the $p-d$ hybridization hoppings. This is physically plausible and indicates how one can simulate the conditions which might be more realistic for material applications. The analysis of correlated models that microscopically include the electron-phonon interaction will be the subject of future studies.
}}

\acknowledgments
W. Brzezicki and A. M. Ole\'s kindly acknowledge the financial support
by National Science Centre (NCN, Poland), under Project
No.~2021/43/B/ST3/02166.
F.~Forte and \mbox{M. Cuoco} kindly acknowledge support by the Project
TOPSPIN funded by the MIUR-PRIN Bando 2017---grant No. 20177SL7HC.
A.~M.~Ole\'s is also grateful for support via the Alexander von
Humboldt Foundation \mbox{Fellowship} \mbox{(Humboldt-Forschungspreis)}.


\appendix

\section{$d^{4}-p^{6}-d^{4}$ exchange in presence of octahedral rotation}
\label{sec:app44}

\begin{table*}
\caption{Spin-orbital Anderson coefficients for the bended
$d^{4}-p^{6}-d^{4}$ bond.
}
\begin{ruledtabular}
\begin{tabular}{cccc}
$\alpha_{n}^{(4)}$ & $q_{2}q_{2}q_{3}$ & $q_{2}q_{2}q_{4}$ & $q_{2}q_{2}q_{5}$ \\ 
\hline
$\alpha_{1}^{(4)}$ & $0$ & $-\frac{1}{2}\left(t_{1}^{2}+t_{2}^{2}\right){}^{2}$ & $-\frac{1}{2}\left(t_{1}^{2}+t_{2}^{2}\right){}^{2}$\tabularnewline
$\alpha_{2}^{(4)}$ & $\frac{1}{6}\left(2t_{1}^{4}+2t_{2}^{2}t_{1}^{2}+2\left(t_{3}^{2}-t_{4}^{2}\right){}^{2}\right)$ & $\frac{1}{6}\left(t_{1}^{4}+t_{2}^{2}t_{1}^{2}-2\left(t_{3}^{2}-t_{4}^{2}\right){}^{2}\right)$ & $\frac{1}{6}\left(3t_{1}^{4}+3t_{2}^{2}t_{1}^{2}\right)$\tabularnewline
$\alpha_{3}^{(4)}$ & $\frac{1}{6}\left(2t_{2}^{4}+2t_{1}^{2}t_{2}^{2}+2\left(t_{3}^{2}-t_{4}^{2}\right){}^{2}\right)$ & $\frac{1}{6}\left(t_{2}^{4}+t_{1}^{2}t_{2}^{2}-2\left(t_{3}^{2}-t_{4}^{2}\right){}^{2}\right)$ & $\frac{1}{6}\left(3t_{2}^{4}+3t_{1}^{2}t_{2}^{2}\right)$\tabularnewline
$\alpha_{4}^{(4)}$ & $-\frac{2}{3}\left(t_{1}^{4}+\left(t_{3}^{2}-t_{4}^{2}\right){}^{2}\right)$ & $\frac{1}{6}\left(t_{1}^{4}+\left(t_{3}^{2}-t_{4}^{2}\right){}^{2}\right)$ & $-\frac{1}{2}\left(t_{1}^{4}+\left(t_{3}^{2}-t_{4}^{2}\right){}^{2}\right)$\tabularnewline
$\alpha_{5}^{(4)}$ & $-\frac{2}{3}\left(t_{2}^{4}+\left(t_{3}^{2}-t_{4}^{2}\right){}^{2}\right)$ & $\frac{1}{6}\left(t_{2}^{4}+\left(t_{3}^{2}-t_{4}^{2}\right){}^{2}\right)$ & $-\frac{1}{2}\left(t_{2}^{4}+\left(t_{3}^{2}-t_{4}^{2}\right){}^{2}\right)$\tabularnewline
$\alpha_{6}^{(4)}$ & $-\frac{2}{3}\left(t_{1}^{2}t_{2}^{2}+\left(t_{3}^{2}-t_{4}^{2}\right){}^{2}\right)$ & $\frac{1}{6}\left(t_{1}^{2}\ t_{2}^{2}+\left(t_{3}^{2}-t_{4}^{2}\right){}^{2}\right)$ & $-\frac{1}{2}\left(t_{1}^{2}t_{2}^{2}+\left(t_{3}^{2}-t_{4}^{2}\right){}^{2}\right)$\tabularnewline
$\alpha_{7}^{(4)}$ & $\frac{2}{3}t_{1}^{2}t_{2}^{2}$ & $-\frac{1}{6}t_{1}^{2}t_{2}^{2}$ & $\frac{1}{2}\text{\ensuremath{t_{1}^{2}t_{2}^{2}}}$\tabularnewline
$\alpha_{8}^{(4)}$ & $\frac{1}{3}t_{1}t_{2}\left(t_{1}^{2}+t_{2}^{2}\right)$ & $\frac{1}{6}t_{1}t_{2}\left(t_{1}^{2}+t_{2}^{2}\right)$ & $\frac{1}{2}t_{1}t_{2}\left(t_{1}^{2}+t_{2}^{2}\right)$\tabularnewline
$\alpha_{9}^{(4)}$ & $0$ & $-\frac{1}{2}t_{2}^{2}\left(t_{3}^{2}-t_{4}^{2}\right)$ & $\frac{1}{2}t_{2}^{2}\left(t_{3}^{2}-t_{4}^{2}\right)$\tabularnewline
$\alpha_{10}^{(4)}$ & $0$ & $\frac{1}{2}t_{1}^{2}\left(t_{3}^{2}-t_{4}^{2}\right)$ & $-\frac{1}{2}t_{1}^{2}\left(t_{3}^{2}-t_{4}^{2}\right)$\tabularnewline
$\alpha_{11}^{(4)}$ & $-\frac{2}{3}t_{1}^{2}\left(t_{3}^{2}-t_{4}^{2}\right)$ & $-\frac{1}{3}t_{1}^{2}\left(t_{3}^{2}-t_{4}^{2}\right)$ & $0$\tabularnewline
$\alpha_{12}^{(4)}$ & $\frac{2}{3}t_{2}^{2}\left(t_{3}^{2}-t_{4}^{2}\right)$ & $\frac{1}{3}t_{2}^{2}\left(t_{3}^{2}-t_{4}^{2}\right)$ & $0$\tabularnewline
$\alpha_{13}^{(4)}$ & $\frac{2}{3}t_{1}^{3}t_{2}$ & $-\frac{1}{6}t_{1}^{3}t_{2}$ & $\frac{1}{2}t_{1}^{3}t_{2}$\tabularnewline
$\alpha_{14}^{(4)}$ & $\frac{2}{3}t_{1}t_{2}^{3}$ & $-\frac{1}{6}t_{1}t_{2}^{3}$ & $\frac{1}{2}t_{1}t_{2}^{3}$\tabularnewline
$\alpha_{15}^{(4)}$ & $\frac{2}{3}t_{1}t_{2}\left(t_{3}^{2}-t_{4}^{2}\right)$ & $\frac{1}{3}t_{1}t_{2}\left(t_{3}^{2}-t_{4}^{2}\right)$ & $0$\tabularnewline
$\alpha_{16}^{(4)}$ & $0$ & $\frac{1}{2}t_{1}t_{2}\left(t_{3}^{2}-t_{4}^{2}\right)$ & $-\frac{1}{2}t_{1}t_{2}\left(t_{3}^{2}-t_{4}^{2}\right)$\tabularnewline
\end{tabular}
\end{ruledtabular}
\label{tab:I}
\end{table*}

\begin{table*}
\caption{Orbital Anderson coefficients for the bended $d^{4}-p^{6}-d^{4}$ bond.}
\begin{ruledtabular}
\begin{tabular}{cccc}
$\beta_{n}^{(4)}$ & $q_{2}q_{2}q_{3}$ & $q_{2}q_{2}q_{4}$ & $q_{2}q_{2}q_{5}$\tabularnewline
\hline
$\beta_{1}^{(4)}$ & $0$ & $\frac{1}{2}\left(t_{1}^{2}+t_{2}^{2}\right){}^{2}$ & $\frac{1}{2}\left(t_{1}^{2}+t_{2}^{2}\right){}^{2}$\tabularnewline
$\beta_{2}^{(4)}$ & $\frac{1}{6}\left(4t_{1}^{4}+4t_{2}^{2}t_{1}^{2}+4\left(t_{3}^{2}-t_{4}^{2}\right){}^{2}\right)$ & $\frac{1}{6}\left(-t_{1}^{4}-t_{2}^{2}t_{1}^{2}+2\left(t_{3}^{2}-t_{4}^{2}\right){}^{2}\right)$ & $\frac{1}{6}\left(-3t_{1}^{4}-3t_{2}^{2}t_{1}^{2}\right)$\tabularnewline
$\beta_{3}^{(4)}$ & $\frac{1}{6}\left(4t_{2}^{4}+4t_{1}^{2}t_{2}^{2}+4\left(t_{3}^{2}-t_{4}^{2}\right){}^{2}\right)$ & $\frac{1}{6}\left(-t_{2}^{4}-t_{1}^{2}t_{2}^{2}+2\left(t_{3}^{2}-t_{4}^{2}\right){}^{2}\right)$ & $\frac{1}{6}\left(-3t_{2}^{4}-3t_{1}^{2}t_{2}^{2}\right)$\tabularnewline
$\beta_{4}^{(4)}$ & $-\frac{4}{3}\left(t_{1}^{4}+\left(t_{3}^{2}-t_{4}^{2}\right){}^{2}\right)$ & $\frac{1}{6}\left(-t_{1}^{4}-\left(t_{3}^{2}-t_{4}^{2}\right){}^{2}\right)$ & $\frac{1}{2}\left(t_{1}^{4}+\left(t_{3}^{2}-t_{4}^{2}\right){}^{2}\right)$\tabularnewline
$\beta_{5}^{(4)}$ & $-\frac{4}{3}\left(t_{2}^{4}+\left(t_{3}^{2}-t_{4}^{2}\right){}^{2}\right)$ & $\frac{1}{6}\left(-t_{2}^{4}-\left(t_{3}^{2}-t_{4}^{2}\right){}^{2}\right)$ & $\frac{1}{2}\left(t_{2}^{4}+\left(t_{3}^{2}-t_{4}^{2}\right){}^{2}\right)$\tabularnewline
$\beta_{6}^{(4)}$ & $-\frac{4}{3}\left(t_{1}^{2}\ t_{2}^{2}+\left(t_{3}^{2}-t_{4}^{2}\right){}^{2}\right)$ & $\frac{1}{6}\left(-t_{1}^{2}t_{2}^{2}-\left(t_{3}^{2}-t_{4}^{2}\right){}^{2}\right)$ & $\frac{1}{2}\left(t_{1}^{2}t_{2}^{2}+\left(t_{3}^{2}-t_{4}^{2}\right){}^{2}\right)$\tabularnewline
$\beta_{7}^{(4)}$ & $\frac{4}{3}t_{1}^{2}t_{2}^{2}$ & $\frac{1}{6}t_{1}^{2}t_{2}^{2}$ & $-\frac{1}{2}t_{1}^{2}t_{2}^{2}$\tabularnewline
$\beta_{8}^{(4)}$ & $\frac{2}{3}t_{1}t_{2}\left(t_{1}^{2}+t_{2}^{2}\right)$ & $-\frac{1}{6}t_{1}t_{2}\left(t_{1}^{2}+t_{2}^{2}\right)$ & $-\frac{1}{2}t_{1}t_{2}\left(t_{1}^{2}+t_{2}^{2}\right)$\tabularnewline
$\beta_{9}^{(4)}$ & $0$ & $\frac{1}{2}t_{2}^{2}\left(t_{3}^{2}-t_{4}^{2}\right)$ & $-\frac{1}{2}t_{2}^{2}\left(t_{3}^{2}-t_{4}^{2}\right)$\tabularnewline
$\beta_{10}^{(4)}$ & $0$ & $-\frac{1}{2}t_{1}^{2}\left(t_{3}^{2}-t_{4}^{2}\right)$ & $\frac{1}{2}t_{1}^{2}\left(t_{3}^{2}-t_{4}^{2}\right)$\tabularnewline
$\beta_{11}^{(4)}$ & $-\frac{4}{3}t_{1}^{2}\left(t_{3}^{2}-t_{4}^{2}\right)$ & $\frac{1}{3}t_{1}^{2}\left(t_{3}^{2}-t_{4}^{2}\right)$ & $0$\tabularnewline
$\beta_{12}^{(4)}$ & $\frac{4}{3}t_{2}^{2}\left(t_{3}^{2}-t_{4}^{2}\right)$ & $-\frac{1}{3}t_{2}^{2}\left(t_{3}^{2}-t_{4}^{2}\right)$ & $0$\tabularnewline
$\beta_{13}^{(4)}$ & $\frac{4}{3}t_{1}^{3}t_{2}$ & $\frac{1}{6}t_{1}^{3}t_{2}$ & $-\frac{1}{2}t_{1}^{3}t_{2}$\tabularnewline
$\beta_{14}^{(4)}$ & $\frac{4}{3}t_{1}t_{2}^{3}$ & $\frac{1}{6}t_{1}t_{2}^{3}$ & $-\frac{1}{2}t_{1}t_{2}^{3}$\tabularnewline
$\beta_{15}^{(4)}$ & $\frac{4}{3}t_{1}t_{2}\left(t_{3}^{2}-t_{4}^{2}\right)$ & $-\frac{1}{3}t_{1}t_{2}\left(t_{3}^{2}-t_{4}^{2}\right)$ & $0$\tabularnewline
$\beta_{16}^{(4)}$ & $0$ & $-\frac{1}{2}t_{1}t_{2}\left(t_{3}^{2}-t_{4}^{2}\right)$ & $\frac{1}{2}t_{1}t_{2}\left(t_{3}^{2}-t_{4}^{2}\right)$\tabularnewline
\end{tabular}
\end{ruledtabular}
\label{tab:II}
\end{table*}

The perturbative expansion involving direct $p-d$ exchange, as well as
Anderson and Goodenough $d-p-d$ superexchange, yields six different
virtual excitation energies given by $-q_{i}^{-1}$ expressed as:
\begin{eqnarray}
q_{1}^{-1} & = & -8U+14J_{H}+2\epsilon_{p},\\
q_{2}^{-1} & = & -4U+7J_{H}+\epsilon_{p},\nonumber\\
q_{3}^{-1} & = & -U+3J_{H},\nonumber\\
q_{4}^{-1} & = & -U,\nonumber\\
q_{5}^{-1} & = & -U-2J_{H}\nonumber,
\end{eqnarray}
where $U$ and $J_{H}$ are Hubbard and Hund's couplings at the metal
ions and $\epsilon_{p}$ is on-site energy of the oxygen orbital.
The hoppings between metal and oxygen can be described by the matrices
\begin{equation}
T_{dp}^{TM_{1}-O}=\!
\begin{pmatrix}0 & 0 & t_{1}\\
0 & 0 & t_{2}\\
t_{3} & t_{4} & 0
\end{pmatrix},\quad
 T_{dp}^{TM_{2}-0}=\!
\begin{pmatrix}0 & 0 & t_{1}\\
0 & 0 & -t_{2}\\
t_{3} & -t_{4} & 0
\end{pmatrix},
\end{equation}
where $T_{dp}^{M_{1}-O}$ describes hopping between the first TM ion and
oxygen and $T_{dp}^{M_{2}-O}$ between second transition metal ion and
oxygen in presence of in-plane octahedral rotation. The rows of these
matrices refer to the $\{d_{yz},d_{zx},d_{xy}\}$ orbital states of the
TM ion and columns to $\{p_{x},p_{y},p_{z}\}$ orbital states of the
oxygen. The hopping amplitudes are given by Slater-Koster rules, i.e.,
\begin{eqnarray}
t_{1} & = & V_{pd\pi}\sin\theta,\nonumber\\
t_{2} & = & V_{pd\pi}\cos\theta,\nonumber\\
t_{3} & = & \! \left[V_{pd\pi}+\left(
-2V_{pd\pi}+\sqrt{3}V_{pd\sigma}\right)\cos^{2}\theta\right]\sin\theta,\nonumber\\
t_{4} & = & \!\left[V_{pd\pi}+\left(
-2V_{pd\pi}+\sqrt{3}V_{pd\sigma}\right)\sin^{2}\theta\right]\cos\theta,
\label{eq:Koster}
\end{eqnarray}
where $\theta$ is the angle of the bond between metal and oxygen
with respect to $\langle 100\rangle$ direction in the $(001)$ plane.
To derive exchange we use a second-order perturbative expansion. In
the atomic ground state we have two spins $S=1$ and orbitals $L=1$
at every metal ion. Oxygen is fully occupied and yields no spin or
orbital degree of freedom. In the second order perturbation expansion
with respect to the hopping, we get the
$d_{1(2)}^{4}p^{6}\leftrightarrows d_{1(2)}^{5}p^{5}$
exchange is given by the Hamiltonian,
\begin{eqnarray}
\label{Hij2}
{\cal H}_{i,j}^{(2)}&=&\beta_{1}^{(2)}
+\beta_{2}^{(2)}\!\left(a_{i}^{\dagger}b_{i}-a_{j}^{\dagger}b_{j}+H.c.\right)
\nonumber\\
&+&\beta_{3}^{(2)}\!\left(a_{i}^{\dagger}a_{i}+a_{j}^{\dagger}a_{j}\right)
 + \beta_{4}^{(2)}\!\left(b_{i}^{\dagger}b_{i}+b_{j}^{\dagger}b_{j}\right).
\end{eqnarray}

\begin{widetext}

Here the coefficients in Eq. \eqref{Hij2} take the following form,
\begin{eqnarray}
\beta_{1}^{(2)} & = & 2\left(t_{1}^{2}+t_{2}^{2}\right)q_{2},\nonumber\\
\beta_{2}^{(2)} & = & -t_{1}t_{2}q_{2},\nonumber\\
\beta_{3}^{(2)} & = & \left(-t_{1}^{2}+t_{3}^{2}+t_{4}^{2}\right)q_{2},\nonumber\\
\beta_{4}^{(2)} & = & \left(-t_{2}^{2}+t_{3}^{2}+t_{4}^{2}\right)q_{2}.
\end{eqnarray}
In the fourth order perturbation expansion with respect to the hopping
we have, inter alia, the Anderson processes where the electron from
the second metal goes through oxygen to the first one and back again,
i.e., $d_{1}^{4}p^{6}d_{2}^{4}\leftrightarrows d_{1}^{5}p^{5}d_{2}^{4}
\leftrightarrows d_{1}^{5}p^{6}d_{2}^{3}$ and analogical process
in the other direction. Eventually one finds a Hamiltonian,

\begin{eqnarray}
{\cal H}_{i,j}^{(4)} & \!=\! & \left(\vec{S}_{i}\!\cdot\!\vec{S}_{j}\right)\!\left\{ \alpha_{1}^{(4)}\!+\!\alpha_{2}^{(4)}\left(a_{i}^{\dagger}a_{i}\!
+\!a_{j}^{\dagger}a_{j}\right)\!+\!\alpha_{3}^{(4)}\left(b_{i}^{\dagger}b_{i}\!
+\!b_{j}^{\dagger}b_{j}\right)\!
+\!\alpha_{4}^{(4)}a_{i}^{\dagger}a_{i}a_{j}^{\dagger}a_{j}\!
+\!\alpha_{5}^{(4)}b_{i}^{\dagger}b_{i}b_{j}^{\dagger}b_{j}\!
+\!\alpha_{6}^{(4)}\left(a_{i}^{\dagger}a_{i}b_{j}^{\dagger}b_{j}\!
+\!a_{j}^{\dagger}a_{j}b_{i}^{\dagger}b_{i}\right)\right.
\nonumber\\
& \!+\! & \alpha_{7}^{(4)}\left(a_{i}^{\dagger}b_{i}\!
+\!b_{i}^{\dagger}a_{i}\right)\!\left(a_{j}^{\dagger}b_{j}\!
+\!b_{j}^{\dagger}a_{j}\right)+\left[\alpha_{8}^{(4)}\left(a_{i}^{\dagger}b_{i}\!
-\!a_{j}^{\dagger}b_{j}\right)\!
+\!\alpha_{9}^{(4)}b_{i}^{\dagger}c_{i}b_{j}^{\dagger}c_{j}\!
+\!\alpha_{10}^{(4)}c_{i}^{\dagger}a_{i}c_{j}^{\dagger}a_{j}\!
+\!\alpha_{11}^{(4)}a_{i}^{\dagger}c_{i}c_{j}^{\dagger}a_{j}\!
+\!\alpha_{12}^{(4)}c_{i}^{\dagger}b_{i}b_{j}^{\dagger}c_{j}\right.
\nonumber\\
& \!+\! & \alpha_{13}^{(4)}\left(a_{i}^{\dagger}a_{i}a_{j}^{\dagger}b_{j}\!
-\!a_{j}^{\dagger}a_{j}a_{i}^{\dagger}b_{i}\right)\!
+\!\alpha_{14}^{(4)}\left(b_{i}^{\dagger}b_{i}a_{j}^{\dagger}b_{j}\!
-\!b_{j}^{\dagger}b_{j}a_{i}^{\dagger}b_{i}\right)\!
+\!\alpha_{15}^{(4)}\left(a_{i}^{\dagger}c_{i}c_{j}^{\dagger}b_{j}\!
-\!a_{j}^{\dagger}c_{j}c_{i}^{\dagger}b_{i}\right)\nonumber\\
& \!+\! & \left.\left.
\!\alpha_{16}^{(4)}\left(c_{i}^{\dagger}b_{i}c_{j}^{\dagger}a_{j}\!
-\!c_{j}^{\dagger}b_{j}c_{i}^{\dagger}a_{i}\right)\!+\!H.c.\right]\right\}
\nonumber \\ & \!+\! & \beta_{1}^{(4)}\!
+\!\beta_{2}^{(4)}\left(a_{i}^{\dagger}a_{i}+a_{j}^{\dagger}a_{j}\right)\!
+\!\beta_{3}^{(4)}\left(b_{i}^{\dagger}b_{i}\!+\!b_{j}^{\dagger}b_{j}\right)\!
+\!\beta_{4}^{(4)}a_{i}^{\dagger}a_{i}a_{j}^{\dagger}a_{j}\!
+\!\beta_{5}^{(4)}b_{i}^{\dagger}b_{i}b_{j}^{\dagger}b_{j}\!
+\!\beta_{6}^{(4)}\left(a_{i}^{\dagger}a_{i}b_{j}^{\dagger}b_{j}\!
+\!a_{j}^{\dagger}a_{j}b_{i}^{\dagger}b_{i}\right)
\nonumber\\ & \!+\! & \beta_{7}^{(4)}\left(a_{i}^{\dagger}b_{i}\!
+\!b_{i}^{\dagger}a_{i}\right)\!\left(a_{j}^{\dagger}b_{j}\!
+\!b_{j}^{\dagger}a_{j}\right)\!
+\!\left[\beta_{8}^{(4)}\left(a_{i}^{\dagger}b_{i}\!
-\!a_{j}^{\dagger}b_{j}\right)\!
+\!\beta_{9}^{(4)}b_{i}^{\dagger}c_{i}b_{j}^{\dagger}c_{j}\!
+\!\beta_{10}^{(4)}c_{i}^{\dagger}a_{i}c_{j}^{\dagger}a_{j}\!
+\!\beta_{11}^{(4)}a_{i}^{\dagger}c_{i}c_{j}^{\dagger}a_{j}\!
+\!\beta_{12}^{(4)}c_{i}^{\dagger}b_{i}b_{j}^{\dagger}c_{j}\right.
\nonumber\\ & \!+\! & \beta_{13}^{(4)}\left(a_{i}^{\dagger}a_{i}a_{j}^{\dagger}b_{j}\!
-\!a_{j}^{\dagger}a_{j}a_{i}^{\dagger}b_{i}\right)\!
+\!\beta_{14}^{(4)}\left(b_{i}^{\dagger}b_{i}a_{j}^{\dagger}b_{j}\!
-\!b_{j}^{\dagger}b_{j}a_{i}^{\dagger}b_{i}\right)\!
+\!\beta_{15}^{(4)}\left(a_{i}^{\dagger}c_{i}c_{j}^{\dagger}b_{j}\!
-\!a_{j}^{\dagger}c_{j}c_{i}^{\dagger}b_{i}\right)
\nonumber\\ & \!+\! & \left.\beta_{16}^{(4)}\left(c_{i}^{\dagger}b_{i}c_{j}^{\dagger}a_{j}\!
-\!c_{j}^{\dagger}b_{j}c_{i}^{\dagger}a_{i}\right)\!+\!H.c.\right],
\end{eqnarray}
where the coefficients are given in Tables I and II. The convention is
such that for example \hfill\break
$\alpha_{1}^{(4)}=
-\frac{1}{2}\left(t_{1}^{2}+t_{2}^{2}\right){}^{2}q_{2}q_{2}q_{4}
-\frac{1}{2}\left(t_{1}^{2}+t_{2}^{2}\right){}^{2}q_{2}q_{2}q_{5}$.
In the fourth-order perturbation expansion, we also have Goodenough
processes where one electron from the oxygen goes to one metal ion
and another electron to the other one, i.e.,
\mbox{$d_{1}^{4}p^{6}d_{2}^{4}\leftrightarrows d_{1}^{5}p^{5}d_{2}^{4}
\leftrightarrows d_{1}^{5}p^{4}d_{2}^{5}$}
an analogical process in the other direction.
This leads to a Hamiltonian,
\begin{eqnarray}
\tilde{{\cal H}}_{i,j}^{(4)} & \!=\! & \left(\vec{S}_{i}\!\cdot\!\vec{S}_{j}\right)\left\{ \tilde{\alpha}_{1}^{(4)}\!+\!\tilde{\alpha}_{2}\left(a_{i}^{\dagger}a_{i}\!
+\!a_{j}^{\dagger}a_{j}\right)\!
+\!\tilde{\alpha}_{3}^{(4)}\left(b_{i}^{\dagger}b_{i}\!
+\!b_{j}^{\dagger}b_{j}\right)\!
+\!\tilde{\alpha}_{4}^{(4)}a_{i}^{\dagger}a_{i}a_{j}^{\dagger}a_{j}\!
+\!\tilde{\alpha}_{5}^{(4)}b_{i}^{\dagger}b_{i}b_{j}^{\dagger}b_{j}\!
+\!\tilde{\alpha}_{6}^{(4)}\left(a_{i}^{\dagger}a_{i}b_{j}^{\dagger}b_{j}\!
+\!a_{j}^{\dagger}a_{j}b_{i}^{\dagger}b_{i}\right)\right.
\nonumber \\ & \!+\! & \tilde{\alpha}_{7}^{(4)}\left(a_{i}^{\dagger}b_{i}\!
+\!b_{i}^{\dagger}a_{i}\right)\!\left(a_{j}^{\dagger}b_{j}\!
+\!b_{j}^{\dagger}a_{j}\right)\!
+\!\left[\tilde{\alpha}_{8}^{(4)}\left(a_{i}^{\dagger}b_{i}\!
-\!a_{j}^{\dagger}b_{j}\right)\!
+\!\tilde{\alpha}_{9}^{(4)}a_{i}^{\dagger}c_{i}c_{j}^{\dagger}a_{j}\!
+\!\tilde{\alpha}_{10}^{(4)}c_{i}^{\dagger}b_{i}b_{j}^{\dagger}c_{j}\right.
\nonumber \\ & \!+\! & \left.\left.\tilde{\alpha}_{11}^{(4)}\left(a_{i}^{\dagger}a_{i}a_{j}^{\dagger}b_{j}\!
-\!a_{j}^{\dagger}a_{j}a_{i}^{\dagger}b_{i}\right)\!
+\!\tilde{\alpha}_{12}^{(4)}\left(b_{i}^{\dagger}b_{i}a_{j}^{\dagger}b_{j}\!
-\!b_{j}^{\dagger}b_{j}a_{i}^{\dagger}b_{i}\right)\!
+\!\tilde{\alpha}_{13}^{(4)}\left(a_{i}^{\dagger}c_{i}c_{j}^{\dagger}b_{j}\!
-\!a_{j}^{\dagger}c_{j}c_{i}^{\dagger}b_{i}\right)\!+\!H.c.\right]\right\}
\nonumber  \\ & \!+\! & \tilde{\beta}_{1}^{(4)}\!+\!\tilde{\beta}_{2}^{(4)}\left(a_{i}^{\dagger}a_{i}\!
+\!a_{j}^{\dagger}a_{j}\right)\!
+\!\tilde{\beta}_{3}^{(4)}\left(b_{i}^{\dagger}b_{i}\!
+\!b_{j}^{\dagger}b_{j}\right)\!
+\!\tilde{\beta}_{4}^{(4)}a_{i}^{\dagger}a_{i}a_{j}^{\dagger}a_{j}\!
+\!\tilde{\beta}_{5}^{(4)}b_{i}^{\dagger}b_{i}b_{j}^{\dagger}b_{j}\!
+\!\tilde{\beta}_{6}^{(4)}\left(a_{i}^{\dagger}a_{i}b_{j}^{\dagger}b_{j}\!
+\!a_{j}^{\dagger}a_{j}b_{i}^{\dagger}b_{i}\right)
\nonumber \\ & \!+\! & \tilde{\beta}_{7}^{(4)}\left(a_{i}^{\dagger}b_{i}\!
+\!b_{i}^{\dagger}a_{i}\right)\!\left(a_{j}^{\dagger}b_{j}\!
+\!b_{j}^{\dagger}a_{j}\right)\!
+\!\left[\tilde{\beta}_{8}^{(4)}\left(a_{i}^{\dagger}b_{i}\!
-\!a_{j}^{\dagger}b_{j}\right)\!
+\!\tilde{\beta}_{9}^{(4)}a_{i}^{\dagger}c_{i}c_{j}^{\dagger}a_{j}\!
+\!\tilde{\beta}_{10}^{(4)}c_{i}^{\dagger}b_{i}b_{j}^{\dagger}c_{j}\right.
\nonumber \\ & \!+\! & \left.\tilde{\beta}_{11}^{(4)}\left(a_{i}^{\dagger}a_{i}a_{j}^{\dagger}b_{j}\!
-\!a_{j}^{\dagger}a_{j}a_{i}^{\dagger}b_{i}\right)\!
+\!\tilde{\beta}_{12}^{(4)}\left(b_{i}^{\dagger}b_{i}a_{j}^{\dagger}b_{j}\!
-\!b_{j}^{\dagger}b_{j}a_{i}^{\dagger}b_{i}\right)\!
+\!\tilde{\beta}_{13}^{(4)}\left(a_{i}^{\dagger}c_{i}c_{j}^{\dagger}b_{j}\!
-\!a_{j}^{\dagger}c_{j}c_{i}^{\dagger}b_{i}\right)+H.c.\right],
\end{eqnarray}
where the coefficients are given in Tabs. III and IV. The convention is
such that for example,
\mbox{$\alpha_{1}^{(4)}=
-2\left(t_{1}^{2}+t_{2}^{2}\right){}^{2}q_{1}q_{2}q_{2}$}.
Then the full spin-orbital exchange Hamiltonian can be composed as
\begin{equation}
{\cal H}_{i,j}\left(\theta\right)=
{\cal H}_{i,j}^{(2)}+{\cal H}_{i,j}^{(4)}+\tilde{{\cal H}}_{i,j}^{(4)},
\end{equation}
being the function of the TM$_i-$O$-$TM$_j$ bond angle $\theta$
encoded in the hopping matrices $T_{d-p}^{{\bf TM}_i-{\bf O}}$ and
$T_{d-p}^{{\bf TM}_j-{\bf O}}$. Note that because these matrices are not
the same \mbox{${\cal H}_{i,j}\left(\theta\right)\not={\cal H}_{j,i}\left(\theta\right)$}.
\end{widetext}

\begin{table}
\caption{Spin-orbital Goodenough coefficients for the bended
$d^{4}-p^{6}-d^{4}$ bond.}
\begin{ruledtabular}
\begin{tabular}{cc}
$\tilde{\alpha}_{n}^{(4)}$ & $q_{1}q_{2}q_{2}$ \tabularnewline
\hline
$\tilde{\alpha}_{1}^{(4)}$ & $-2\left(t_{1}^{2}+t_{2}^{2}\right){}^{2}$ \tabularnewline
$\tilde{\alpha}_{2}^{(4)}$ & $2t_{1}^{2}\left(t_{1}^{2}+t_{2}^{2}\right)$ \tabularnewline
$\tilde{\alpha}_{3}^{(4)}$ & $2t_{2}^{2}\left(t_{1}^{2}+t_{2}^{2}\right)$ \tabularnewline
$\tilde{\alpha}_{4}^{(4)}$ & $-2\left(t_{1}^{4}+\left(t_{3}^{2}-t_{4}^{2}\right){}^{2}\right)$  \tabularnewline
$\tilde{\alpha}_{5}^{(4)}$ & $-2\left(t_{2}^{4}+\left(t_{3}^{2}-t_{4}^{2}\right){}^{2}\right)$  \tabularnewline
$\tilde{\alpha}_{6}^{(4)}$ & $-2\left(t_{1}^{2}t_{2}^{2}+\left(t_{3}^{2}-t_{4}^{2}\right){}^{2}\right)$  \tabularnewline
$\tilde{\alpha}_{7}^{(4)}$ & $2t_{1}^{2}t_{2}^{2}$ \tabularnewline
$\tilde{\alpha}_{8}^{(4)}$ & $2t_{1}t_{2}\left(t_{1}^{2}+t_{2}^{2}\right)$ \tabularnewline
$\tilde{\alpha}_{9}^{(4)}$ & $2t_{1}^{2}\left(t_{4}^{2}-t_{3}^{2}\right)$ \tabularnewline
$\tilde{\alpha}_{10}^{(4)}$ & $2t_{2}^{2}\left(t_{3}^{2}-t_{4}^{2}\right)$ \tabularnewline
$\tilde{\alpha}_{11}^{(4)}$ & $2t_{1}^{3}t_{2}$ \tabularnewline
$\tilde{\alpha}_{12}^{(4)}$ & $2t_{1}t_{2}^{3}$ \tabularnewline
$\tilde{\alpha}_{13}^{(4)}$ & $2t_{1}t_{2}\left(t_{3}^{2}-t_{4}^{2}\right)$ \tabularnewline
\end{tabular}$\quad$%
\end{ruledtabular}
\end{table}

\begin{table}
\caption{Orbital Goodenough coefficients for the bended
$d^{4}-p^{6}-d^{4}$ bond.}
\begin{ruledtabular}
\begin{tabular}{cc}
$\tilde{\beta}_{n}^{(4)}$ & $q_{1}q_{2}q_{2}$\tabularnewline
\hline
$\tilde{\beta}_{1}^{(4)}$ & $2\left(t_{1}^{2}+t_{2}^{2}\right){}^{2}$\tabularnewline
$\tilde{\beta}_{2}^{(4)}$ & $-2\left(t_{1}^{2}+t_{2}^{2}\right)\left(t_{1}^{2}-2\ \left(t_{3}^{2}+t_{4}^{2}\right)\right)$\tabularnewline
$\tilde{\beta}_{3}^{(4)}$ & $-2\left(t_{1}^{2}+t_{2}^{2}\right)\left(t_{2}^{2}-2\ \left(t_{3}^{2}+t_{4}^{2}\right)\right)$\tabularnewline
$\tilde{\beta}_{4}^{(4)}$ & $2\left(t_{1}^{4}-4\left(t_{3}^{2}+t_{4}^{2}\right)t_{1}^{2}+t_{3}^{4}
+t_{4}^{4}+6t_{3}^{2}\ t_{4}^{2}\right)$\tabularnewline
$\tilde{\beta}_{5}^{(4)}$ & $2\left(t_{2}^{4}-4\left(t_{3}^{2}+t_{4}^{2}\right)t_{2}^{2}+t_{3}^{4}+t_{4}^{4}+6\ t_{3}^{2}t_{4}^{2}\right)$\tabularnewline
$\tilde{\beta}_{6}^{(4)}$ & $2\left(t_{3}^{4}+6t_{4}^{2}t_{3}^{2}+t_{1}^{2}t_{2}^{2}+t_{4}^{4}
-2\left(t_{1}^{2}+t_{2}^{2}\right)\left(t_{3}^{2}+t_{4}^{2}\right)\right)$
\tabularnewline
$\tilde{\beta}_{7}^{(4)}$ & $-2t_{1}^{2}t_{2}^{2}$\tabularnewline
$\tilde{\beta}_{8}^{(4)}$ & $-2t_{1}t_{2}\left(t_{1}^{2}+t_{2}^{2}\right)$\tabularnewline
$\tilde{\beta}_{9}^{(4)}$ & $2t_{1}^{2}\left(t_{4}^{2}-t_{3}^{2}\right)$\tabularnewline
$\tilde{\beta}_{10}^{(4)}$ & $2t_{2}^{2}\left(t_{3}^{2}-t_{4}^{2}\right)$\tabularnewline
$\tilde{\beta}_{11}^{(4)}$ & $-2t_{1}t_{2}\left(t_{1}^{2}-2\left(t_{3}^{2}+t_{4}^{2}\right)\right)$
\tabularnewline
$\tilde{\beta}_{12}^{(4)}$ & $-2t_{1}t_{2}\left(t_{2}^{2}-2\left(t_{3}^{2}+t_{4}^{2}\right)\right)$
\tabularnewline
$\tilde{\beta}_{13}^{(4)}$ & $2t_{1}t_{2}\left(t_{3}^{2}-t_{4}^{2}\right)$\tabularnewline
\end{tabular}$\quad$%
\end{ruledtabular}
\end{table}
\newpage

\begin{widetext}

\section{$d^{3}-p^{6}-d^{4}$ exchange in presence of octahedral rotation}
\label{sec:app34}

The perturbative expansion involving direct $p-d$ exchange, including
Anderson and Goodenough $d-p-d$ superexchange yields seven different
virtual excitation energies given by $-q_{i}^{-1}$ expressed as:
\begin{eqnarray}
q_{1}^{-1} & = & 4\tilde{J}_{H}-3\tilde{U}+\epsilon_{p},\nonumber\\
q_{2}^{-1} & = & -I_{e}+7J_{H}-4U+\epsilon_{p},\nonumber\\
q_{3}^{-1} & = & -I_{e}+4\tilde{J}_{H}+7J_{H}-3\tilde{U}-4U+2\epsilon_{p},\nonumber\\
q_{4}^{-1} & = & -I_{e}+4\tilde{J}_{H}-4J_{H}-3\tilde{U}+3U,\nonumber\\
q_{5}^{-1} & = & -I_{e}+4\tilde{J}_{H}-7J_{H}-3\tilde{U}+3U,\nonumber\\
q_{6}^{-1} & = & -I_{e}+4\tilde{J}_{H}-9J_{H}-3\tilde{U}+3U,\nonumber\\
q_{7}^{-1} & = & -I_{e}-6\tilde{J}_{H}+7J_{H}+2\tilde{U}-4U,
\end{eqnarray}
where $U$ and $J_{H}$ are Hubbard and Hund's couplings at the $d^{4}$
metal ions, $\tilde{U}$ and $\tilde{J}_{H}$ are for the $d^{3}$ ions,
and $\epsilon_{p}$ is onsite energy of the oxygen.

The hoppings between TM and oxygen site can be described by the matrices,
\begin{equation}
T_{d-p}^{{\bf M}_{i}-{\bf O}}=\begin{pmatrix}0 & 0 & \tilde{t}_{1}\\
0 & 0 & \tilde{t}_{2}\\
\tilde{t}_{3} & \tilde{t}_{4} & 0
\end{pmatrix},\qquad \qquad
T_{d-p}^{{\bf M}_{j}-{\bf O}}=\begin{pmatrix}0 & 0 & t_{1}\\
0 & 0 & -t_{2}\\
t_{3} & -t_{4} & 0
\end{pmatrix},
\end{equation}
where $T_{d-p}^{{\bf M}_{i}-{\bf O}}$ describes hopping between first
($d^3$) metal ion and oxygen and $T_{d-p}^{{\bf M}_j-{\bf O}}$ between
second metal ($d^{4}$) ion and oxygen in presence of in-plane
octahedral rotation. The rows of these matrices refer to $d_{yz}$,
$d_{zx}$, and $d_{xy}$ orbital states of the metal,
and columns to $p_{x}$, $p_{y}$, and $p_{z}$ orbital states of the
oxygen. The hopping amplitudes are given by Slater-Koster rules as
before, see Eq. \ref{eq:Koster}, where the $\pi$ and $\sigma$ bonding
amplitudes are a priori different for the $d^3-p^6$ and $d^4-p^6$ bonds.

To derive exchange we use a second-order perturbative expansion. In the
atomic ground state we have spin $S=1$ and orbital $L=1$ at the $d^{4}$
metal ion and spin $S=3/2$ and orbital $L=0$ at the $d^{3}$ metal ion.
Oxygen is fully occupied and yields no spin or orbital degree of freedom.
In the second order perturbation expansion with respect to the hopping,
we get the $d_{1}^{3}p^{6}\leftrightarrows d_{1}^{4}p^{5}$ and $d_{2}^{4}p^{6}\leftrightarrows d_{2}^{5}p^{5}$ exchanges given by the
Hamiltonian,
\begin{equation}
{\cal H}_{i,j}^{(2)}=
\beta_{1}^{(2)}+\beta_{2}^{(2)}\left(a_{j}^{\dagger}b_{j}+H.c.\right)
+\beta_{3}^{(2)}a_{j}^{\dagger}a_{j}+\beta_{4}^{(2)}b_{j}^{\dagger}b_{j},
\end{equation}
where the coefficients take the following form,
\begin{eqnarray}
\beta_{1}^{(2)} & = & \left(\tilde{t}_{1}^{2}+\tilde{t}_{2}^{2}+\tilde{t}_{3}^{2}
+\tilde{t}_{4}^{2}\right)q_{1}+\left(t_{1}^{2}+t_{2}^{2}\right)q_{2},
\nonumber\\
\beta_{2}^{(2)} & = & t_{1}t_{2}q_{2},\nonumber\\
\beta_{3}^{(2)} & = & \left(t_{3}^{2}+t_{4}^{2}-t_{1}^{2}\right)q_{2},\nonumber\\
\beta_{4}^{(2)} & = & \left(t_{3}^{2}+t_{4}^{2}-t_{2}^{2}\right)q_{2}.
\end{eqnarray}
In the fourth order perturbation expansion with respect to the hopping
we have, inter alia, the Anderson processes where the electron from
the second metal goes through oxygen to the first one and back again,
i.e., $d_{1}^{3}p^{6}d_{2}^{4}\leftrightarrows d_{1}^{4}p^{5}d_{2}^{4}
\leftrightarrows d_{1}^{4}p^{6}d_{2}^{3}$
and $d_{1}^{3}p^{6}d_{2}^{4}\leftrightarrows d_{1}^{3}p^{5}d_{2}^{5}
\leftrightarrows d_{1}^{2}p^{6}d_{2}^{5}$. This leads to a Hamiltonian,
%
\begin{eqnarray}
{\cal H}_{i,j}^{(4)} & \!=\! & \left(\vec{S}_{i}\!\cdot\!\vec{S}_{j}\right)\!
 \times\!\left\{\! \alpha_{1}^{(4)}\!+\!\alpha_{2}^{(4)}\!\left(
 a_{j}^{\dagger}b_{j}\!+\!H.c.\right)
\!+\!\alpha_{3}^{(4)}a_{j}^{\dagger}a_{j}
\!+\!\alpha_{4}^{(4)}b_{j}^{\dagger}b_{j}\!\right\}
\!+\!\beta_{1}^{(4)}\!+\!\beta_{2}^{(4)}\!\left(\!a_{j}^{\dagger}b_{j}
 +H.c.\!\right)\!+\!\beta_{3}^{(4)}\!a_{j}^{\dagger}a_{j}
\!+\!\beta_{4}^{(4)}\!b_{j}^{\dagger}b_{j},
\end{eqnarray}
where the coefficients take the following form,
\begin{eqnarray}
\alpha_{1}^{(4)} & = & \tfrac{1}{18}\left[-6\left(t_{1}^{2}+t_{2}^{2}\right)\left(
\tilde{t}_{1}^{2}+\tilde{t}_{2}^{2}\right)q_{2}^{2}q_{7}-3\left(t_{1}^{2}
+t_{2}^{2}\right)\left(\tilde{t}_{1}^{2}+\tilde{t}_{2}^{2}\right)q_{1}^{2}
\left(q_{5}+q_{6}\right)+4\left(t_{3}\tilde{t}_{3}
-t_{4}\tilde{t}_{4}\right)^{2}q_{1}^{2}
\left(q_{4}-q_{5}\right)\right],\nonumber \\
\alpha_{2}^{(4)} & = & -\tfrac{1}{18}t_{1}t_{2}\left(\tilde{t}_{1}^{2}+\tilde{t}_{2}^{2}\right)
\left(q_{1}^{2}\left(4q_{4}-q_{5}+3q_{6}\right)+6q_{2}^{2}q_{7}\right),
\nonumber \\ \alpha_{3}^{(4)} & = & \tfrac{1}{18}\left[]t_{1}^{2}\left(\tilde{t}_{1}^{2}+\tilde{t}_{2}^{2}\right)
-\left(t_{3}\tilde{t}_{3}-t_{4}\tilde{t}_{4}\right)^{2}\right]\left(q_{1}^{2}
\left(4q_{4}-q_{5}+3q_{6}\right)+6q_{2}^{2}q_{7}\right),\nonumber \\
\alpha_{4}^{(4)} & = & \tfrac{1}{18}\left[]t_{2}^{2}\left(\tilde{t}_{1}^{2}+\tilde{t}_{2}^{2}\right)
-\left(t_{3}\tilde{t}_{3}-t_{4}\tilde{t}_{4}\right)^{2}\right]\left(q_{1}^{2}
\left(4q_{4}-q_{5}+3q_{6}\right)+6q_{2}^{2}q_{7}\right), 
\end{eqnarray}
and
\begin{eqnarray}
\beta_{1}^{(4)} & = & \tfrac{1}{12}\left[6\left(t_{1}^{2}+t_{2}^{2}\right)\left(\tilde{t}_{1}^{2}
+\tilde{t}_{2}^{2}\right)q_{2}^{2}q_{7}+3\left(t_{1}^{2}+t_{2}^{2}\right)
\left(\tilde{t}_{1}^{2}+\tilde{t}_{2}^{2}\right)q_{1}^{2}
\left(q_{5}+q_{6}\right)+4\left(t_{3}\tilde{t}_{3}
-t_{4}\tilde{t}_{4}\right)^{2}q_{1}^{2}\left(2q_{4}+q_{5}\right)\right],
\nonumber \\ \beta_{2}^{(4)} & = & -\tfrac{1}{12}t_{1}t_{2}\left(\tilde{t}_{1}^{2}+\tilde{t}_{2}^{2}\right)
\left[q_{1}^{2}\left(8q_{4}+q_{5}-3q_{6}\right)-6q_{2}^{2}q_{7}\right],
\nonumber \\ \beta_{3}^{(4)} & = &
 \tfrac{1}{12}\left[t_{1}^{2}\left(\tilde{t}_{1}^{2}+\tilde{t}_{2}^{2}\right)
-\left(t_{3}\tilde{t}_{3}-t_{4}\tilde{t}_{4}\right)^{2}\right]\left(q_{1}^{2}
\left(8q_{4}+q_{5}-3q_{6}\right)-6q_{2}^{2}q_{7}\right),\nonumber \\
\beta_{4}^{(4)} & = & \tfrac{1}{12}\left[t_{2}^{2}\left(\tilde{t}_{1}^{2}+\tilde{t}_{2}^{2}\right)
-\left(t_{3}\tilde{t}_{3}-t_{4}\tilde{t}_{4}\right)^{2}\right]
\left(q_{1}^{2}\left(8q_{4}+q_{5}-3q_{6}\right)-6q_{2}^{2}q_{7}\right).
\end{eqnarray}
In the fourth-order perturbation expansion, we also have Goodenough
processes where one electron from the oxygen goes to one metal and
another electron to the other one, i.e.,
$d_{1}^{3}p^{6}d_{2}^{4}\leftrightarrows d_{1}^{4}p^{5}d_{2}^{4}
\leftrightarrows d_{1}^{4}p^{4}d_{2}^{5}$, and
\mbox{$d_{1}^{3}p^{6}d_{2}^{4}\leftrightarrows d_{1}^{3}p^{5}d_{2}^{5}
\leftrightarrows d_{1}^{4}p^{4}d_{2}^{5}$.}
This leads to a Hamiltonian,
\begin{eqnarray}
{\cal\tilde{H}}_{i,j}^{(4)}&\!=\!&
\left(\vec{S}_i\!\cdot\!\vec{S}_j\right)\!\times\!\left\{ \tilde{\alpha}_{1}^{(4)}\!+\!\tilde{\alpha}_{2}^{(4)}\!\left(\!a_{j}^{\dagger}b_{j}\!
+\!H.c.\!\right)\!+\!\tilde{\alpha}_{3}^{(4)}a_{j}^{\dagger}a_{j}
\!+\!\tilde{\alpha}_{4}^{(4)}b_{j}^{\dagger}b_{j}\right\}\!
+\!\tilde{\beta}_{1}^{(4)}\!+\!\tilde{\beta}_{2}^{(4)}\!\left(\!a_{j}^{\dagger}b_{j}
+H.c.\!\right)\!
+\!\tilde{\beta}_{3}^{(4)}a_{j}^{\dagger}a_{j}\!
+\!\tilde{\beta}_{4}^{(4)}b_{j}^{\dagger}b_{j},  
\end{eqnarray}
%
where the coefficients take the following form,
\begin{eqnarray}
\tilde{\alpha}_{1}^{(4)} & = & -\tfrac{1}{3}\left(t_{1}^{2}+t_{2}^{2}\right)\left(\tilde{t}_{1}^{2}
+\tilde{t}_{2}^{2}\right)q_{3}\left(q_{1}+q_{2}\right)^{2}, \nonumber \\
\tilde{\alpha}_{2}^{(4)} & = & -\tfrac{1}{3}t_{1}t_{2}\left(\tilde{t}_{1}^{2}+\tilde{t}_{2}^{2}\right)q_{3}
\left(q_{1}+q_{2}\right)^{2},\nonumber \\
\tilde{\alpha}_{3}^{(4)} & = & \tfrac{1}{3}\left[t_{1}^{2}\left(\tilde{t}_{1}^{2}+\tilde{t}_{2}^{2}\right)
-\left(t_{3}\tilde{t}_{3}-t_{4}\tilde{t}_{4}\right)^{2}\right]q_{3}
\left(q_{1}+q_{2}\right)^{2},\nonumber \\
\tilde{\alpha}_{4}^{(4)} & = & \tfrac{1}{3}\left[t_{2}^{2}\left(\tilde{t}_{1}^{2}
+\tilde{t}_{2}^{2}\right)-\left(t_{3}\tilde{t}_{3}
-t_{4}\tilde{t}_{4}\right)^{2}\right]q_{3}
\left(q_{1}+q_{2}\right)^{2}, \nonumber \\
\tilde{\beta}_{1}^{(4)} & = & \frac{1}{2}\left(t_{1}^{2}+t_{2}^{2}\right)\left(\tilde{t}_{1}^{2}+\tilde{t}_{2}^{2}
+2\tilde{t}_{3}^{2}+2\tilde{t}_{4}^{2}\right)q_{3}\left(q_{1}+q_{2}\right)^{2},
\nonumber \\ \tilde{\beta}_{2}^{(4)} & = & \frac{1}{2}t_{1}t_{2}\left(\tilde{t}_{1}^{2}+\tilde{t}_{2}^{2}+2\tilde{t}_{3}^{2}
+2\tilde{t}_{4}^{2}\right)q_{3}\left(q_{1}+q_{2}\right)^{2},
\nonumber \\
\tilde{\beta}_{3}^{(4)} & = & -\frac{1}{2}q_{3}\left(q_{1}+q_{2}\right)^{2}\left[\left(\tilde{t}_{1}^{2}
+\tilde{t}_{2}^{2}\right)\left(t_{1}^{2}-2\tilde{t}_{3}^{2}
-2\tilde{t}_{4}^{2}\right)
\left(2t_{1}^{2}-t_{3}^{2}-t_{4}^{2}\right)\left(\tilde{t}_{3}^{2}
+\tilde{t}_{4}^{2}\right)-\left(t_{3}\tilde{t}_{4}
+t_{4}\tilde{t}_{3}\right)^{2}\right], \nonumber \\
\tilde{\beta}_{4}^{(4)} & = & -\frac{1}{2}q_{3}\left(q_{1}+q_{2}\right)^{2}\left[\left(\tilde{t}_{1}^{2}
+\tilde{t}_{2}^{2}\right)\left(t_{2}^{2}-2\tilde{t}_{3}^{2}
-2\tilde{t}_{4}^{2}\right)
+\left(2t_{2}^{2}-t_{3}^{2}-t_{4}^{2}\right)\left(\tilde{t}_{3}^{2}
+\tilde{t}_{4}^{2}\right)-\left(t_{3}\tilde{t}_{4}
+t_{4}\tilde{t}_{3}\right)^{2}\right].
\end{eqnarray}
Then the full spin-orbital exchange Hamiltonian for the impurity bond
can be composed as
\begin{equation}
{\cal H}_{i,j}^{\rm imp}\left(\theta\right)=
{\cal H}_{i,j}^{(2)}+{\cal H}_{i,j}^{(4)}+\tilde{{\cal H}}_{i,j}^{(4)},
\end{equation}
being the function of the TM$_i-{\rm O}-$TM$_j$ bond angle $\theta$,
encoded in the hopping matrices $T_{d-p}^{{\rm TM}_i-{\rm O}}$, and
$T_{d-p}^{{\rm TM}_j-{\rm O}}$. Here also obviously assume that
${\cal H}_{i,j}\left(\theta\right)\not={\cal H}_{j,i}\left(\theta\right)$.
\end{widetext}

\section{Role of $J_H$ and $\lambda$ in a $d^4$ plaquette within th effective spin-orbital model}
\label{sec:app_renorm}

{\color{black}{
In this Appendix we present the results for the $d^4$ plaquette as obtained by means of effective spin-orbital Hamiltonian where Hund's exchange and spin-orbit coupling are scaled up by a factor of $3.2$. The factor $\chi = 3.2$ is such that the $J_H/U$ ratio has the same amplitude of that used in the single bond multi-orbital Hubbard model.
Indeed, the ratio of the values of the Hubbard $U$ interaction used in the effective spin-orbital exchange approach with respect to the $U$ used in the multi-orbital Hubbard model calculations is 3.2 ($U=8$ eV in the effective spin-orbital model vs. $U=2.5$ eV in the multi-orbital Hubbard model). The results without magnetic order are shown in Fig. \ref{fig:11} and with a $C$-AF magnetic order in Fig. \ref{fig:12}. The outcomes indicate that the qualitative behavior does not change and it is substantially the same as that described in Sec. \ref{sec:44}. However, the energy windows of $V_{pd\pi}$ hybridization amplitude and temperature where the optimal angles profiles exhibit signatures of variations, are different. }}

\begin{figure}[t!]
\includegraphics[width=1\columnwidth]{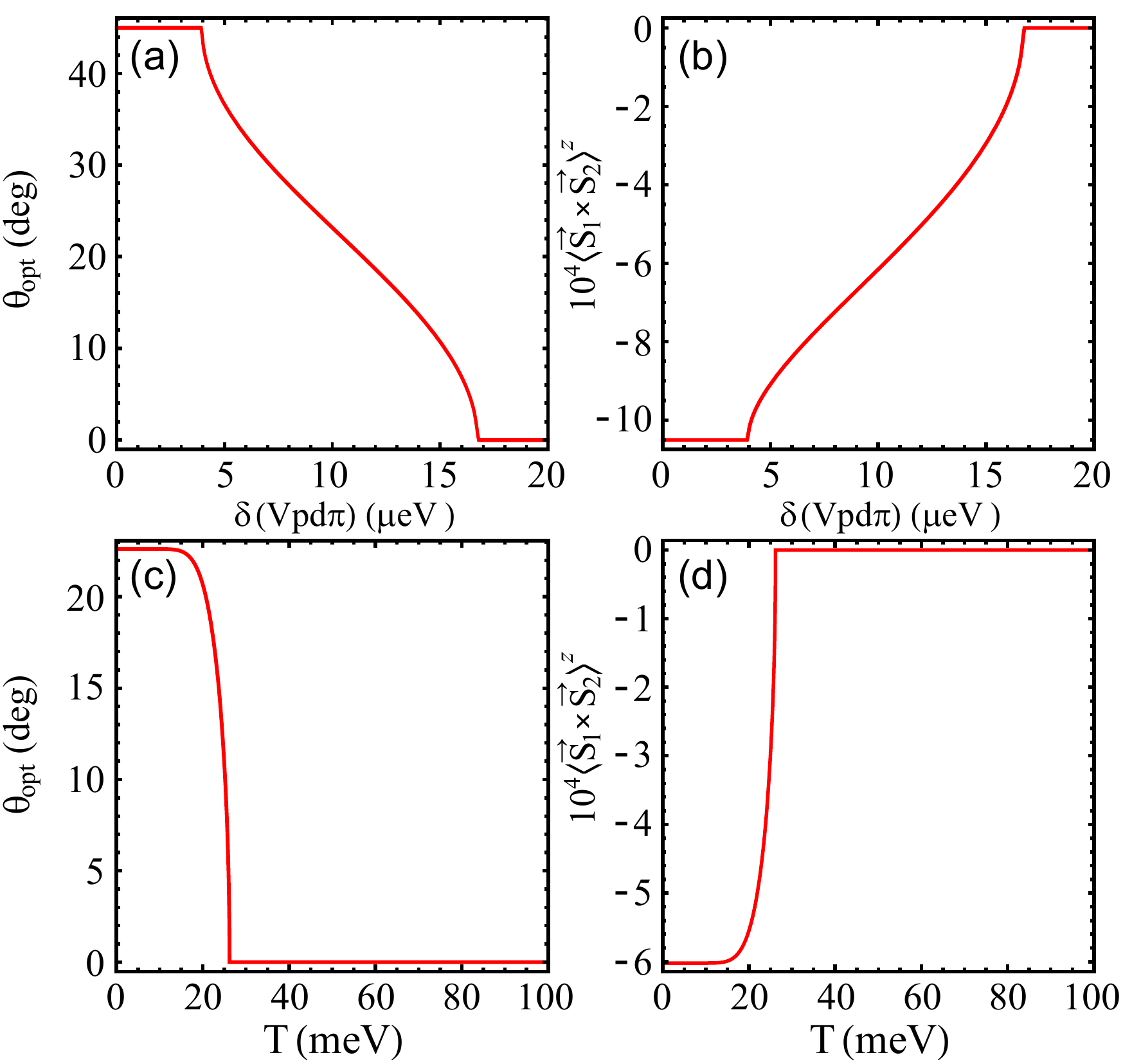}
\caption{
Results for the undoped $d^4$ plaquette with finite $J_H$ and $\lambda$ increased by a factor of $\chi = 3.2$ compared to Sec. \ref{sec:44}:
(a,b) typical curve of the optimal bond angle and the ground state
average of $\langle\vec{S}_1\times\vec{S}_2\rangle^z$ versus
$\delta(V_{pd\pi})\equiv V_{pd\pi}-V_{0}$ for $V_{pd\sigma}=1.5$ eV in
the intermediate-angle phase;
(c,d) thermal dependencies of the optimal bond angle and average
$(\vec{S}_1\times\vec{S}_2)^z$ in the intermediate phase for
$V_{pd\sigma}=1.5$ and $V_{pd\pi}=1.2962$ eV.
The other parameters are: $U=8.0$, $J_H=1.6$, $\epsilon_p=-4.5$,
$\delta=0.35$, $\delta_{ort}=0.09$, and $\lambda = 0.24$, all in eV.
\label{fig:11}}
\end{figure}

\begin{figure}[t!]
\includegraphics[width=1\columnwidth]{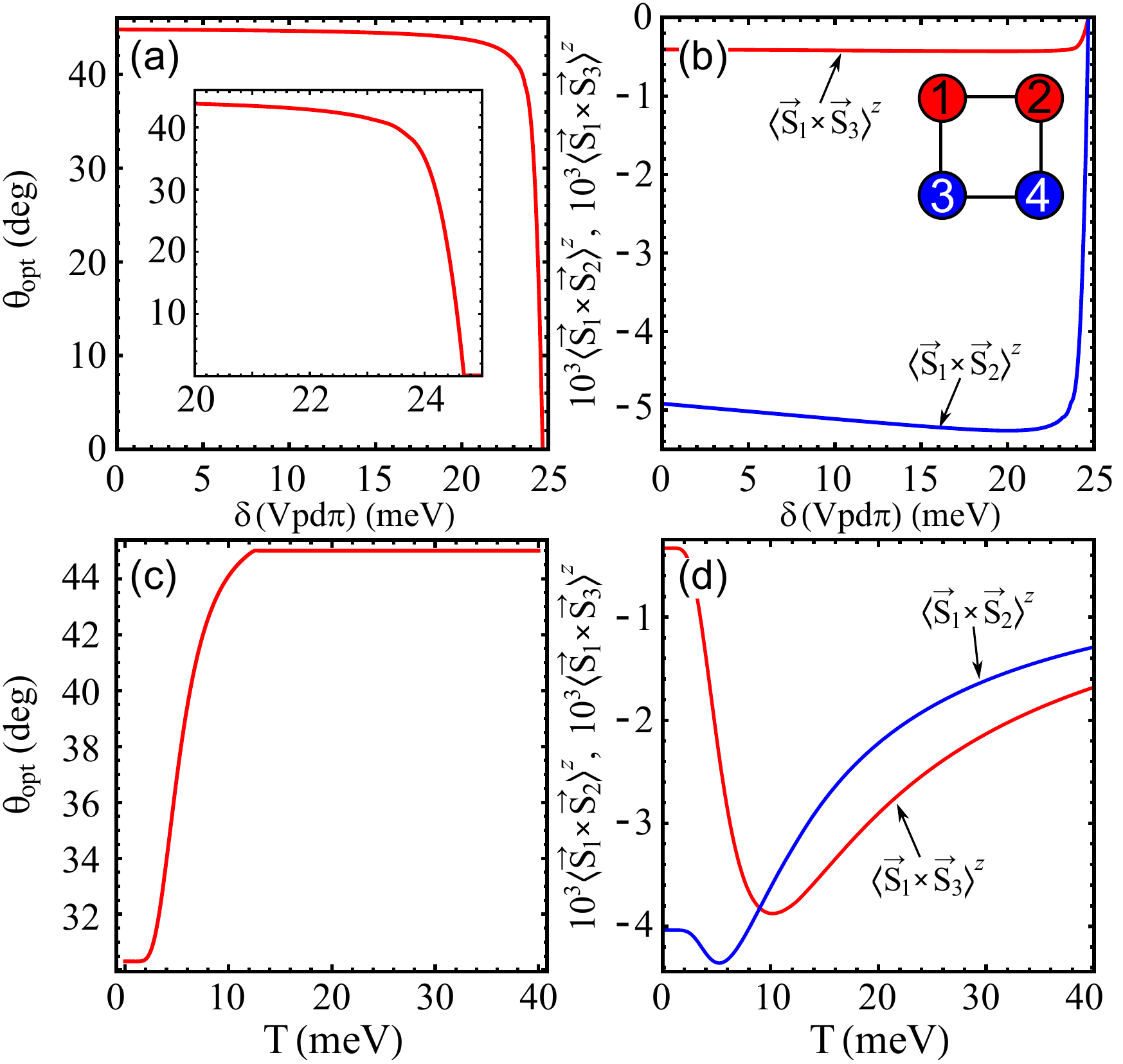}
\caption{
Results for the undoped $d^4$ plaquette with $C$-AF magnetic order (see the inset) and with $J_H$ and $\lambda$ increased by a factor of $\chi = 3.2$ compared to Sec. \ref{sec:44}:
(a,b) typical curves of the optimal bond angle and the ground state
averages of $\langle\vec{S}_1\times\vec{S}_2\rangle^z$ and
$\langle\vec{S}_1\times\vec{S}_3\rangle^z$ versus
$\delta(V_{pd\pi})=V_{pd\pi}-V_{0}$ for $V_{pd\sigma}=1.5$ eV in the
intermediate phase between $\theta=0$ and $\theta=\pi/4$;
(c,d)~thermal dependencies of the optimal bond angle and average
$\langle\vec{S}_1\times\vec{S}_2\rangle^z$ and
$\langle\vec{S}_1\times\vec{S}_3\rangle^z$ in the intermediate phase for
$V_{pd\sigma}=1.5$ and $V_{pd\pi}=1.2928$ eV. The other parameters are:
$U=8.0$, $J_H=1.6$, $\epsilon_p=-4.5$, $\delta=0.35$,
$\delta_{ort}=0.09$, $\lambda=0.24$ and $h=0.2$, all in eV.
\label{fig:12}}
\end{figure}


\section{Single bond within the effective spin-orbital exchange approach}
\label{sec:1bond_eff}

{\color{black}{
In Fig. \ref{fig:13} we show the temperature-dependent curves of the optimal angles for single $d^4-d^4$ and $d^3-d^4$ bonds.
The results should be compared with Sec. \ref{sec:bond}. We see that the trends observed for a full Hubbard model for large $U$ are reproduced in this effective approach, i.e., the angle drops with temperature and lower Hund's exchange gives larger bond angles. }}

\begin{figure}[t!]
\includegraphics[width=1\columnwidth]{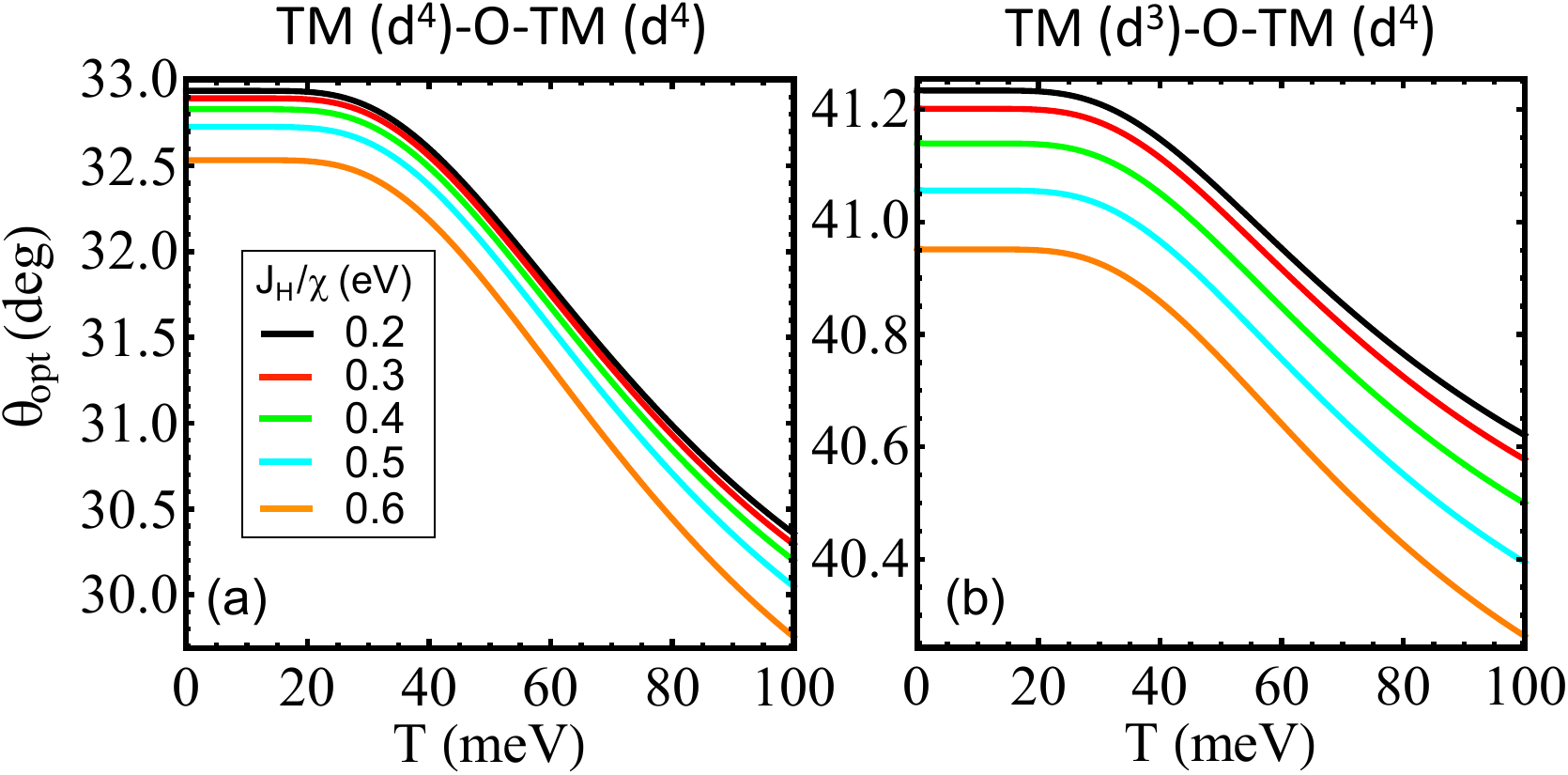}
\caption{
Optimal angle versus temperature for a single: (a) $d^4$---$d^4$ bond and (b) $d^3$---$d^4$ bond, for different values of $J_H$ (renormalized by a factor of $\chi = 3.2$).
The other parameters are:
$V_{pd\sigma}=1.6$, $V_{pd\pi}=1.3$,
$U=8.0$, $\epsilon_p=-4.5$, $\delta=0.25$,
$\delta_{ort}=0.09$, $\lambda=0.24$ and $h=0.2$, all in eV.
\label{fig:13}}
\end{figure}

\section{Role of the lattice potential: a phenomenological approach}
\label{sec:latt}

{In the absence of the lattice response, the free energy analysis of the electronic model leads to solutions of the optimal bond-angles that correspond to extreme rotations of the octahedra and maximal bond angle variations occur in a small range of amplitudes for the $p-d$ hopping parameters. Here we show how this this problem can be overcome by including an effective electron-lattice potential.
To this aim, we construct a phenomenological model that includes an effective electron-lattice potential that is quadratic in the bond angle. We assume that the coupling constant is proportional to the optimal angle that is obtained by minimizing the free energy of the purely electronic system. Such an assumption is physically plausible because the lattice acts as a restoring force and it is reasonable to expect that its strength, through the electron-phonon coupling, can be dependent on the amplitude of the optimal angle that is set by the electronic degrees of freedom. The proposed phenomenological form of the total free energy can be hence expressed as:
\begin{equation}
F(\theta) = F_{{\rm el}}(\theta) + k\,(\theta_{{\rm el}}/\pi)(\theta/\pi)^2,
\label{el-ph}
\end{equation}
where $k$ is an elastic constant.
Therefore, we assume that the lattice deformation potential is harmonic with the stiffness being proportional to $\theta_{{\rm el}}$, the angle that minimizes the electronic part $F_{{\rm el}}(\theta)$.

We have implemented the phenomenological model in Eq. (E1) for the case of the $d^4$ plaquette.
In Fig. \ref{fig:14}(a) we show the typical curves of the electronic free energy at zero temperature as a function of the bond angle. We note that the curves have a single minimum that moves from $0$ to $\pi/4$ in a small window of energy variation for the $V_{pd\pi}$ parameter (of the order of $\mu$ eV). In Fig. \ref{fig:14}(b) we present the optimal angle versus $V_{pd\pi}$ and versus temperature (inset) for the electron-phonon model of Eq. (\ref{el-ph}). We notice that by a proper choice of the constant coupling $k$, which should be of the same order as the electronic free energy change between $0$ and $\pi/4$ away from the transition region, one can stabilize an intermediate-angle phase at a scale of up to tens of eV in $V_{pd\pi}$. Such a phase is demonstrated to exhibit a NTE effect at finite temperature, as it can be seen from an inspection of the inset in Fig. \ref{fig:14}(b)).
}}

\begin{figure}[t!]
\includegraphics[width=0.75\columnwidth]{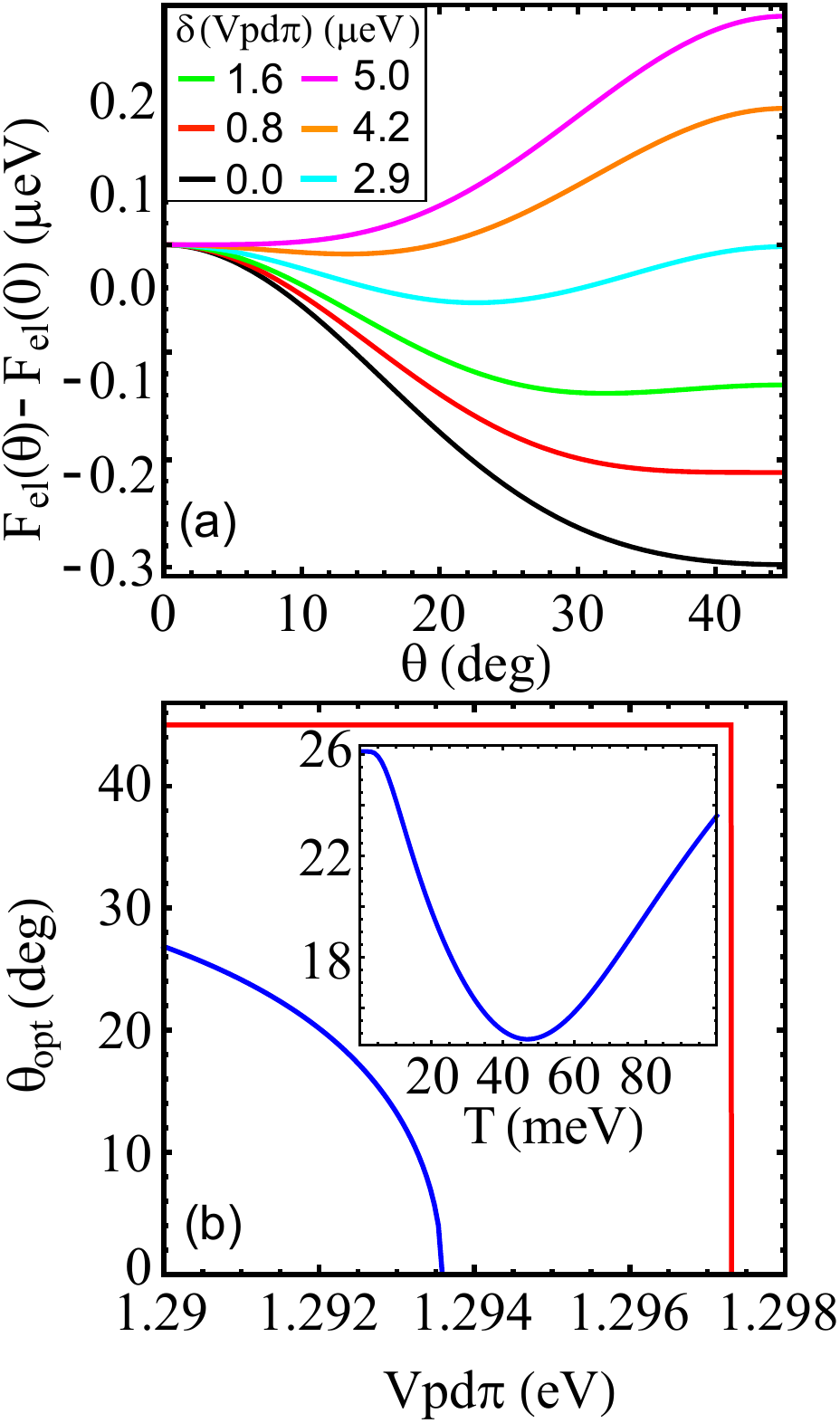}
\caption{
Results for a pure $d^4$ plaquette as obtained within the effective low-energy spin-orbital exchange approach:
(a)~electronic free energy versus bond angle for a given $V_{pd\sigma}\!=1.5$ eV and different values of $V_{pd\pi}\!=\! V_{0}+ \delta(V_{pd\pi})$ chosen in the intermediate-angle phase,
(b) optimal angle versus $V_{pd\pi}$
for a purely electronic free energy (red curve) and
for the phenomenological eletronic-phononic free energy [Eq. (E1)] assuming a representative value for the coupling constant, i.e., $k=6$ meV. The inset of (b) shows the optimal angle versus temperature dependence for $V_{pd\pi}=1.2902$ eV. The other electronic parameters are:
$U=8.0$, $\epsilon_p=-4.5$, $\delta=0.35$,
$\delta_{ort}=0.09$, and $\lambda=0.075$, all values are in eV.
\label{fig:14}}
\end{figure}

\end{document}